\documentclass[10pt]{article}
\usepackage{cite}

\usepackage{graphicx}        % standard LaTeX graphics tool
\usepackage{amssymb}
\usepackage{amsmath}
\usepackage{pdflscape}
\usepackage{tikz}
\usepackage[scriptsize]{subfigure}
\usepackage{amsmath}
\usepackage{soul}
\usepackage{float}

\usepackage[T1]{fontenc}
\usepackage{kpfonts,baskervald}

\usepackage[utf8]{inputenc}
\usepackage{authblk}
\usepackage{hyphenat}
\usepackage{cite}

\usepackage{placeins}

\usepackage{caption} % para poner 'matrices' de figuras
\usepackage{dblfloatfix} % para poner figuras que cubran toda la columna

\tikzset{cross/.style={cross out, draw=black, minimum size=2*(#1-\pgflinewidth), inner sep=0pt, outer sep=0pt},
cross/.default={2.5pt}}
\usepackage[margin=1in]{geometry}
\usepackage{parskip}
\usepackage{titlesec}

\usepackage{authblk} % This package is commonly used for author and affiliation formatting

\usepackage{IEEEtrantools}

\usepackage{cancel}
\usepackage{multirow}

\DeclareGraphicsExtensions{.pdf,.png,.jpg,.eps} %solo para PDFLaTeX
\usepackage{epstopdf}
\usepackage[acronym,shortcuts]{glossaries}

\makeglossaries

\newacronym{GFVSC}{\textcolor{black}{GFM-VSC}}{Grid-forming voltage source converter}

\begin{document}

\title{Impact of current limiters and fast voltage boosters in grid-forming VSC-based generators on transient stability}

 \author{R\'egulo E. \'Avila-Mart\'inez$^1$, Javier Renedo$^2$, Luis Rouco$^1$, Aurelio Garcia-Cerrada$^1$, Lukas Sigrist$^1$, \and Xavier Guillaud$^3$, Taoufik Qoria$^4$}
\date{%
    $^1$ Instituto de Investigación Tecnológica (IIT), ETSI ICAI, Universidad Pontificia Comillas, Madrid, Spain. \\%
    $^2$ ETSI ICAI, Universidad Pontificia Comillas, Madrid, Spain.  \\% 
    $^3$ Laboratoire d’Electrotechniquede et d'Electronique de Puissance (L2EP), Ecole Centrale de Lille, Lille, France. \\
    $^4$ GE Grid Solutions GmbH, Berlin, Germany. \\[2ex]%
    \today
}

\maketitle
\begin{abstract}
Transient stability is a complex phenomenon presented in multi-machine and multi-converter systems, and it is still considered a key limiting factor for stressed power systems. The increasing integration of non-synchronous generation further emphasises the need to address the challenges of improving the transient stability faced by these power systems. Several studies have focused on developing control strategies for \ac{GFVSC} to improve transient stability. These strategies include the use of current limiting algorithms and/or control of active/reactive power injections. This paper investigates the impact of fast voltage boosters (FVBs) and hybrid current limiters (HCLs) on transient stability of power systems with 100\% grid-forming VSC-based generators. Short-circuit simulations and critical clearing time analysis are performed to evaluate the effectiveness of HCLs and FVBs in improving transient stability. The simulation results demonstrate the effectiveness of these approaches in avoiding the loss of synchronism. This research contributes to the current studies on transient stability in power systems and provides valuable insights into the potential of HCLs and FVBs as effective approaches to improve system stability. 
\end{abstract}

This is an unabridged draft of the following paper (submitted and accepted in 23rd Power Systems Computation Conferencen (PSCC'2024), Electric Power Systems Research Journal, special section: VSI:PSCC 2024):
\begin{itemize}
\item R. E. Ávila-Martínez, J. Renedo, L. Rouco, A. García-Cerrada, L. Sigrist, X. Guillaud, T. Qoria, "Impact of current limiters and fast voltage boosters in grid-forming VSC-based generators on transient stability", submitted to 23rd Power System Computation Conference (PSCC), pp. 1-11, 2024.
\item ID: EPSR-D-23-03769R1.
\item Internal reference: IIT-24-108C.
\end{itemize}
%\newpage
%\vspace{-0.7cm}
\section{Introduction}\label{sec:intro}
\noindent Voltage source converters with grid-forming control (\ac{GFVSC}) play an important role in future power systems dominated by Converter-Interfaced Generation (CIG)~\cite{Paolone2020}. \acp{GFVSC} interface renewable energy sources or energy storage systems with electrical power systems by controlling the output voltage magnitude and imposing the VSC frequency at the point of common coupling (PCC). The key feature of the \acp{GFVSC} is that it allows the creation of a grid. Self-synchronisation strategies are needed when operating several \acp{GFVSC} in parallel, in order to ensure that all converters reach the same frequency in steady state \cite{Barker2021}. Self-synchronisation strategies use supplementary controller strategies to mimic the behaviour of synchronous machines, and several variants have been proposed in previous work, such as virtual synchronous machine (VSM) control~\cite{DArco2015,jroldan2019} or Integral-Proportional (IP) control~\cite{Qoria2020}, among others.

Transient stability is a complex phenomenon present in multi-machine and multi-converter systems, and it is considered a key limiting factor in stressed power systems. Previous studies have explored various methods to enhance transient stability in power systems utilizing \acp{GFVSC}, due to their flexible control. Active-power supplementary control strategies in \acp{GFVSC} have been investigated in~\cite{Choopani2020,Xiong2021,Qoria2020,Collados2023}, and voltage/reactive-power supplementary control strategies in \acp{GFVSC} have been proposed in~\cite{XiongX_2021a,RAvilaM2022,BlaabjergFTSAngle2022, SiW2023}. For example, the work in ~\cite{RAvilaM2022} proposed Fast Voltage Boosters (FVBs) in \acp{GFVSC} to improve transient stability, using to different approaches, using local and global measurements. FVBs consists on adding a supplementary voltage set-point in the \ac{GFVSC} during faults and proved to be an effective alternative to improve transient stability. 

Alternatively, current limiters in \acp{GFVSC} can be exploited as shown in~\cite{BoFang,Qoria_VSC_current_limit2020,Qoria_VSC_CCT2020,Rokrok_TS2021}. Although their main objective is to limit the magnitude of the converter's current injection, they can also have a significant impact on loss-of-sycnhronism phenomenon in \acp{GFVSC}, since they limit electrical power provided by the converter. The work in~\cite{Qoria_VSC_current_limit2020} analyses the impact of current-limiting strategies on transient stability of a \ac{GFVSC} connected to an infinite grid. The study analyses two current-limiting strategies: current-modulus limiter in vector control (Current Saturation Algorithm, CSA) and virtual-impedance-based (VI-CL) current limiter. Eventually, the work proposes a combination of both strategies (hybrid current limiter, HCL), improving the performance of the current limitation process as well as transient stability. The work in~\cite{Qoria_VSC_CCT2020} analysed transient stability behaviour of a \ac{GFVSC}, using VI-CLs and proposing an adaptive droop to improve the performance. Meanwhile, the work in~\cite{Rokrok_TS2021} analysed the impact of the angle of the current when using CSAs on transient stability of a \ac{GFVSC} connected to an infinite grid, finding an optimal angle. In~\cite{Guangya2021}, transient stability of \acp{GFVSC} was improved significantly, using the concept of virtual unsaturated power as feedback for the self-synchronisation mechanism of the \ac{GFVSC}, especially when the converter is forced into current limit operation. Further analysis on the performance of the virtual unsaturated power has been presented in~\cite{Laba2023}.

Since~\cite{RAvilaM2022} analysed FVBs in \acp{GFVSC} using CSA-based current limiters only, a question that remains open is whether FVBs maintain their effectiveness on transient stability when using other types of current limiters. In particular, it is of interest to analyse the performance of FVBs when using HCLs, since they improve transient stability of \acp{GFVSC}, in addition to their main application, which is to limit the current of the power converter. Although their different nature, both, FVBs and HCLs, add terms to the voltage set-point of the \ac{GFVSC}. In fact, in some cases, the supplementary voltage set-point provided by the FVB and the HCL may be remarkably different. For example, in case of short circuits close to the \ac{GFVSC}, during the fault, the FVB will include a positive supplementary voltage set-point value, while the HCL will include a negative supplementary voltage set-point value. Hence, it is essential to assess the robustness of FVBs and to analyse potential interactions when using FVBs together with HCLs. 

Along this line, the contributions of this paper are as follows:
\begin{itemize}
	\item Analysis of the impact of FVBs~\cite{RAvilaM2022} and HCLs~\cite{Qoria_VSC_current_limit2020} in \acp{GFVSC} on transient stability in power systems with 100\% CIG. 
	\item Demonstration of the effectiveness of FVBs when implemented with HCLs. Not only no negative interactions were observed, but also the use of FVBs together with HCLs produces significant improvements in the critical clearing times (CCTs) of different faults.
	\item Analysis of the impact of communication latency on the performance of the global control strategy FVB-WACS~\cite{RAvilaM2022}, when using HCLs. 
\end{itemize}

The performance of the control strategies is demonstrated by a theoretical analysis using the Equal Area Criterion (EAC) and by simulation using Kundur's two-area test system~\cite{Kundur1994a} with 100\% grid-forming VSC-based generation. 

\newpage

\section{Grid-forming VSCs}\label{sec:VSC_V}
\subsection{Modelling and control}\label{sec:VSC_V_model}

\begin{figure}[!htbp]
	\begin{center}
		\includegraphics[width=0.7\columnwidth]{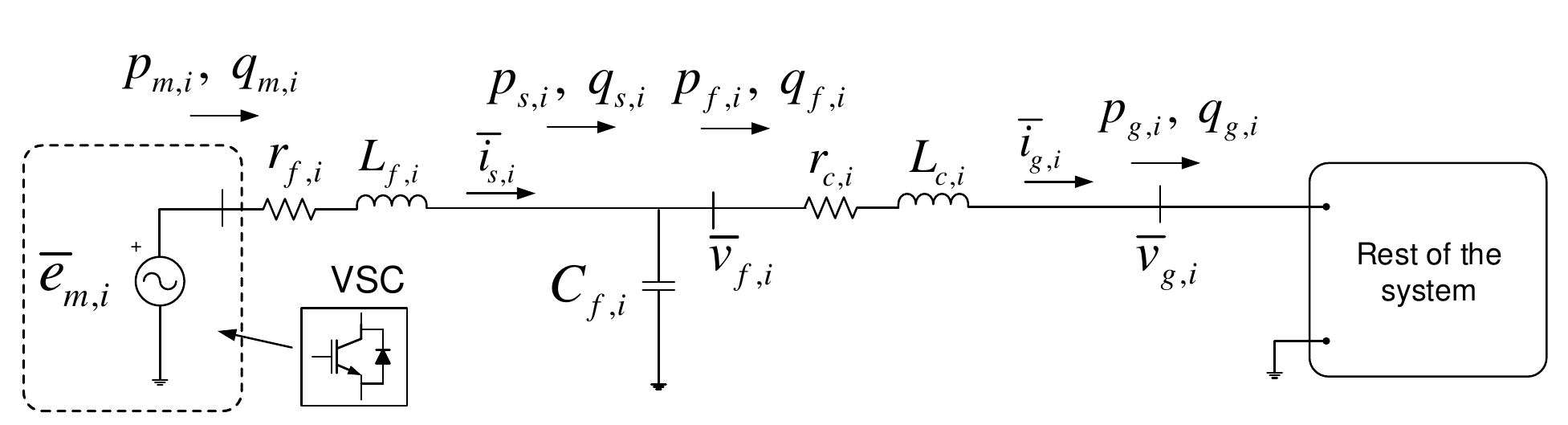}
		\caption{Model of a grid-forming VSC. }
		\label{fig:VSC_V_model}
	\end{center}
\end{figure}

\noindent Fig.~\ref{fig:VSC_V_model} depicts the equivalent model of a \ac{GFVSC}-$i$, which consists of a voltage source~($\bar{e}_{m,i}$) connected to the system through a LC filter and a transformer~\cite{Qoria2020}. 

Fig.~\ref{fig:VSC_V_control_loops_general} presents the \ac{GFVSC}-$i$ general control scheme~\cite{Qoria2020}, which consists of a cascade AC voltage and current control loops in $d-q$ reference frame including: (a) a voltage controller, (b) a virtual transient resistance (used to provide damping during the transient~\cite{Qoria2020}), (c) a current controller (with current limitation), (d) a voltage modulation (with modulation index limitations) and (e) a VSM control used for the self-synchronisation of the \ac{GFVSC}~\cite{Paolone2020,Qoria2020}. 

The \ac{GFVSC}-$i$ controls voltage magnitude $v_{f,i}$ and frequency $\omega_{f,i}$ (and, thus, the voltage angle $\delta_{f,i}$) at bus $f_{i}$. The angle $\delta_{f,i}$ (rad) is imposed by the \ac{GFVSC} VSM control (e) (see Section~\ref{sec:VSC_V_VSM}), aligning $\bar{v}_{f,i}$ with the direct axis component ($d$-axis) of the rotating $d-q$ reference frame.

\FloatBarrier
\subsection{Virtual synchronous machine control (VSM)}\label{sec:VSC_V_VSM}
\begin{figure}[!htbp]
	\begin{center}
 		\vspace{-0.4cm}
		\includegraphics[width=0.85\columnwidth]{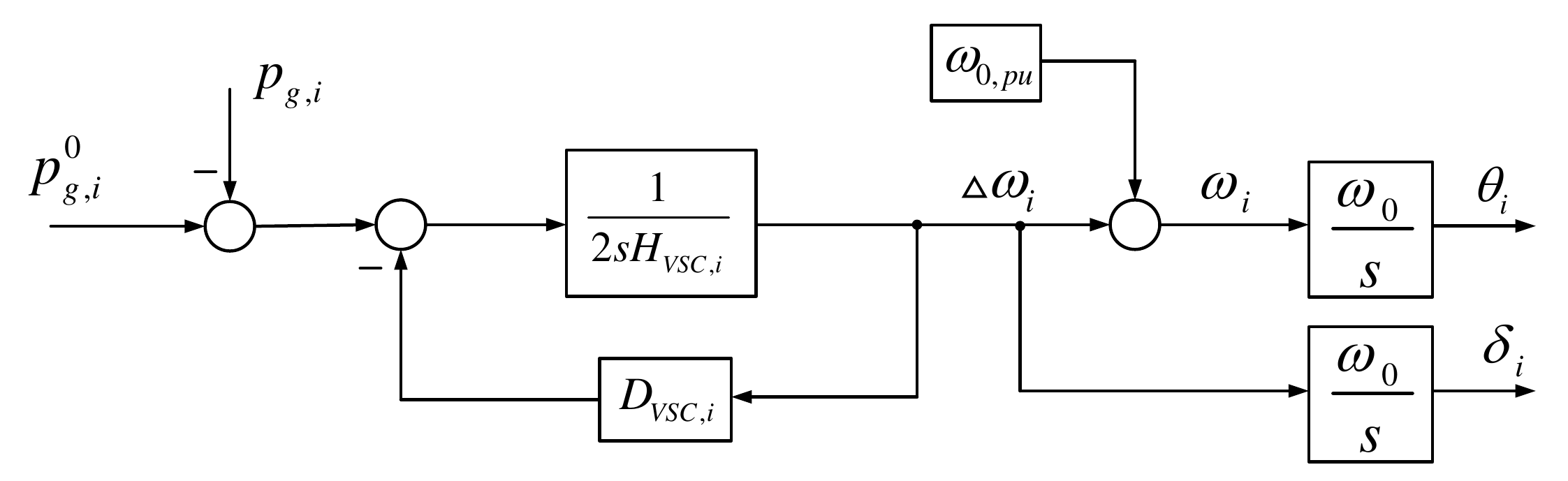}
		\caption{VSM control in a \ac{GFVSC}.}
		\label{fig:VSC_V_VSM}
	\end{center}
\end{figure}

\noindent One option for a self-synchronisation method in \acp{GFVSC} is the so-called virtual synchronous machine (VSM), as shown in Fig.~\ref{fig:VSC_V_VSM}~\cite{DArco2015,Choopani2020}. 

The swing equation emulated by a VSM is given by:
\begin{eqnarray}\label{eq_VSC_V_VSM_v1}
	2 H_{VSC,i}\frac{d \Delta \omega_{i}}{dt} & = & p_{g,i}^{0} - p_{g,i} -D_{VSC,i} \Delta \omega_{i} 
\end{eqnarray}

where:
\begin{itemize}
	\item $H_{VSC,i}$~(s) is the emulated inertia constant.
	\item $D_{VSC,i}$~(pu) is the proportional gain of the primary frequency response (PFR). It also plays the role of damping coefficient. 
	\item $\Delta \omega_{i}=\omega_{i}-\omega_{0,pu}$~(pu), where $\omega_{i}$ is the frequency imposed by the \ac{GFVSC} and $\omega_{0,pu}=1$~pu. 
	\item $p_{g,i}^{0}$~(pu) is a constant active-power set-point of the \ac{GFVSC} at the connection point.
	\item $p_{g,i}$~(pu) is the active-power injected by the \ac{GFVSC} at the connection point.
	\item $\omega_0$ is the nominal frequency in rad/s.
\end{itemize}

\begin{figure*}[!htbp]
	%\centering
	\begin{center}
		\includegraphics[width=1\columnwidth]{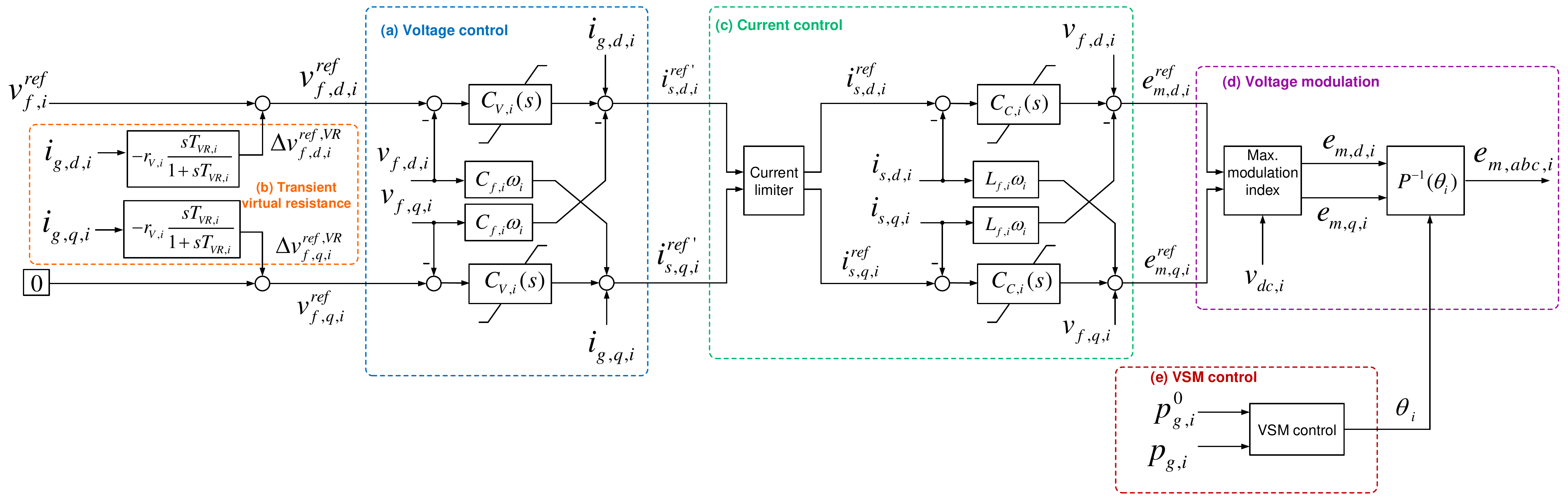}
		\setlength{\belowcaptionskip}{-20pt}
		\caption{General scheme of the control system of a grid-forming VSC. }
		\label{fig:VSC_V_control_loops_general}
	\end{center}
\end{figure*}

 By means of equation~(\ref{eq_VSC_V_VSM_v1}), the \ac{GFVSC}-$i$ imposes the frequency $\omega_{i}$~(pu) at connection point (bus $f_{i}$). The angles in Fig.~\ref{fig:VSC_V_VSM} are angle used for Park's Transform (Fig.~\ref{fig:VSC_V_control_loops_general}).

\section{Current limiters (CL)}\label{sec:VSC_V_current_lim}
\noindent This section briefly reviews the current limiters of \acp{GFVSC} analysed in this work~\cite{Qoria2020}: (a) current saturation algorithm (CSA), (b) current limitation based on virtual impedance (VI-CL) and (c) hybrid current limiter (HCL).

\subsection{Current saturation algorithm (CSA)}\label{sec:VSC_V_current_lim_CSA}
\noindent The current set-points ($i_{s,d,i}^{ref'}$ and $i_{s,q,i}^{ref'}$) of Fig.~\ref{fig:VSC_V_control_loops_general} are the outputs of the outer control loop:
\begin{equation}\label{eq_VSC_GM_inner_isp_dq_ref}
	\bar{i}_{s,i}^{ref'} = i_{s,d,i}^{ref'} +j i_{s,q,i}^{ref'} = i_{s,i}^{ref'} e^{j \delta_{is,i}^{ref'}}
\end{equation}

\begin{figure}[!htbp]
	\begin{center}
		\includegraphics[width=0.30\columnwidth]{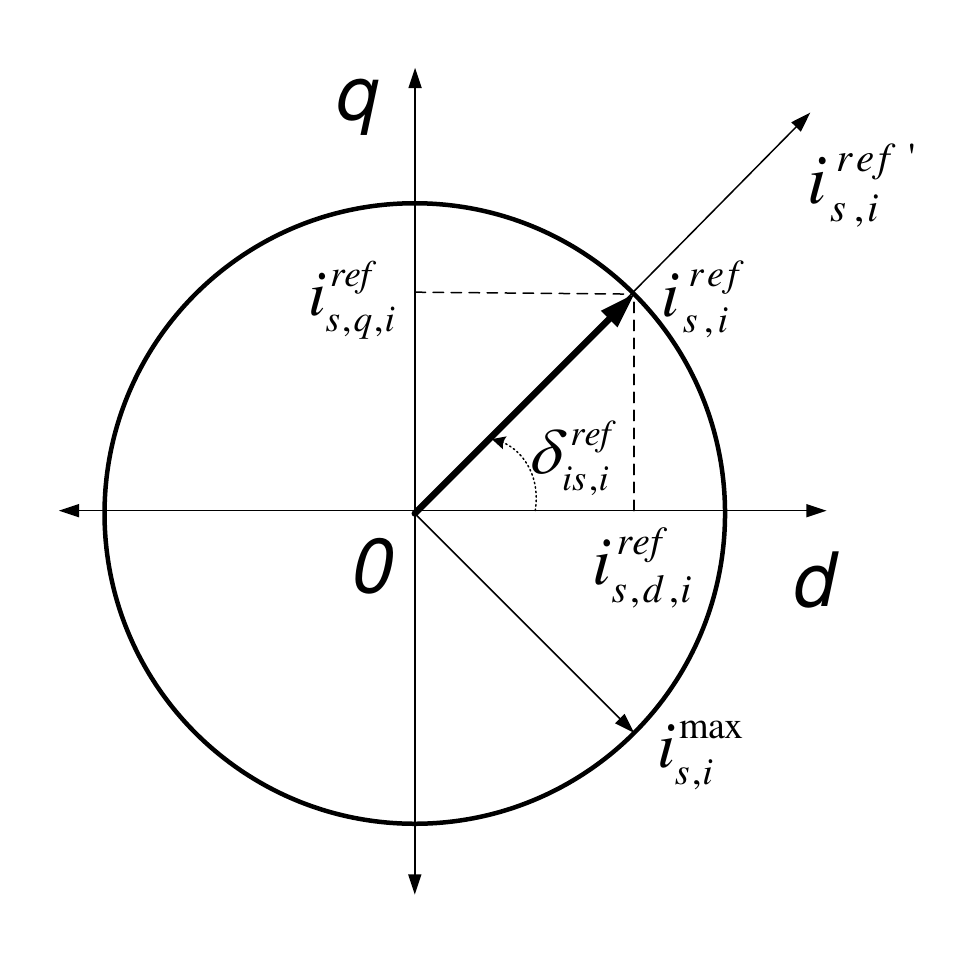}
		\setlength{\belowcaptionskip}{-5pt}
		\caption{Current saturation algorithm (CSA).}
		\label{fig:VSC_V_CSA_diagram}
	\end{center}
\end{figure}

The magnitude of the resulting current set-point, $i_{s,i}^{ref}$, must be lower than or equal to the maximum current of VSC-$i$: $i_{s,i}^{max}$, as illustrated in Fig.~\ref{fig:VSC_V_CSA_diagram}. Typical values of the maximum current are $i_{s,i}^{max}=1-1.25$~pu. Equal priority for $d$- and $q$-axis currents will be used in this work.
\vspace{-0.5cm}

\subsection{Current limitation based on virtual impedance (VI-CL)}\label{sec:VSC_V_current_lim_VI}
\noindent When a fault is detected, a transient virtual impedance is included into the voltage set-point of the \ac{GFVSC} to limit the its current and to improve transient stability \cite{Paquette,Qoria_VSC_CCT2020}:
\begin{eqnarray}\label{eq_VSC_GM_out_Vf_vi}
\begin{array}{cc}
     v_{f,d,i}^{ref} = & v_{f,i}^{ref} + \Delta v_{f,d,i}^{ref,VR} - \Delta v_{f,d,i}^{ref',VI}\\ 

     v_{f,q,i}^{ref} = & 0 + \Delta v_{f,q,i}^{ref,VR} - \Delta v_{f,q,i}^{ref'VI}
\end{array}
\end{eqnarray}
where the terms of the virtual voltage drop read:
\begin{eqnarray}\label{eq_VSC_b_Vf_vi}
\begin{array}{cc}
    \Delta v_{f,d,i}^{ref',VI} = & r_{VI} i_{s,d,i} - x_{VI} i_{s,q,i} \\

    \Delta v_{f,q,i}^{ref',VI} = &  r_{VI} i_{s,q,i} + x_{VI} i_{s,d,i}
\end{array}
\end{eqnarray}
The VI-CL is activated only when the converter current magnitude $i_{s,i}$ exceeds a threshold value ($i_{s,i}>i_{thres,i}^{VI}$). Terms 
$x_{VI}$ and $r_{VI}$ are given in~(\ref{eq_x_vi_Vf}) and (\ref{eq_r_vi_Vf}).
%The expressions of the virtual impedance (XV I and RV I ) are given in (12) and (13).
\begin{eqnarray}\label{eq_x_vi_Vf}
    x_{VI} = \left\lbrace\begin{array}{ccc} 
        k_{p_{r_{VI}}} \sigma_{x/r} \Delta i_{s,i}, & if & i_{s,i}> i_{thres,i}^{VI}
        \\
        0, & if & i_{s,i} \leq i_{thres,i}^{VI} 
        \end{array}\right.
\end{eqnarray}
\begin{equation}\label{eq_r_vi_Vf}
    r_{VI} = x_{VI}/\sigma_{x/r}
\end{equation}
where $\Delta i_{s,i} = i_{s,i} - i_{thres,i}^{VI}$. Parameters $k_{p_{r_{VI}}}$ and $\sigma_{x/r}$, are defined in~\cite{Qoria_VSC_CCT2020}.

\subsection{Hybrid current limiter (HCL)}\label{sec:VSC_V_current_lim_hybrid_CL}
\noindent The work in~\cite{Qoria_VSC_current_limit2020} showed that (a) CSA is more effective than VI-CL to limit the current injection of the \ac{GFVSC}, while (b) VI-CL presentes better behaviour than CSA in terms of transient stability. Hence, reference~\cite{Qoria_VSC_current_limit2020} proposed a hybrid current limiter (HCL), combining CSA and VI-CL current-limitation algorithms to improve the overall performance. In HCLs, both current limiters CSA and VI-CL are implemented simultaneously and their activation is coordinated with thresholds. If $i_{s,i}^{max}$ and $i_{thres,i}^{VI}$ are the maximum current of CSA and the threshold current of VI-CL, respectively, then $i_{thres,i}^{VI}<i_{s,i}^{max}$ must be satisfied. HCL results in an effective algorithm to limit the current, as well as to improve the transient stability behaviour of the \ac{GFVSC}. 

\FloatBarrier
\section{Fast voltage boosters (FVBs) %\color{red}to improve transient stability
}\label{sec:VSC_V_TS}
\noindent Fast voltage boosters (FVBs) for \acp{GFVSC} were proposed in~\cite{RAvilaM2022} to improve transient stability, motivated by the excitation boosters (EBs) used in synchronous machines~\cite{LuisDM2017,LuisDM2019}. FVBs in \acp{GFVSC} are based on reactive-power/voltage control, adding a supplementary voltage set-point, $\Delta v_{f,i}^{ref,TS}$, when faults occur in order to improve transient stability:
\begin{equation}\label{eq:VSC_vf_ref_TS}
	v_{f,i}^{ref} = v_{f,i}^{0} + \Delta v_{f,i}^{ref,TS}
\end{equation}
 where the voltage set-point of the \acp{GFVSC} is the one described in Fig. \ref{fig:VSC_V_control_loops_general}. 

The active power injection of a \ac{GFVSC} (Fig.~\ref{fig:VSC_V_model}) can be approximated as follows: 
\begin{equation}\label{eq:VSC_pg_electrical}
	p_{g,i} \simeq \frac{v_{f,i} v_{g,i}}{x_{c,i}} \sin (\delta_{f,i}-\delta_{g,i})
\end{equation}

When a short circuit occurs, the voltage at the connection point of the \ac{GFVSC}, $v_{g,i}$, decreases. Hence, its active-power injection, $p_{g,i}$, also decreases. According to the emulated swing equation of the \ac{GFVSC} (\ref{eq_VSC_V_VSM_v1}), this results in an acceleration of the converter (i.e., its frequency increases). If the fault is severe enough, the \acp{GFVSC} could eventually lose synchronism. Appropriate control strategies should pull together the frequencies of \acp{GFVSC} to avoid a loss of synchronism between them. A FVB manipulates the voltage at the connection point of the \ac{GFVSC}, $v_{f,i}$ in order to improve transient stability of the system. By changing $v_{f,i}$ in \acp{GFVSC}, the active power injection changes $p_{g,i}$, which is related to their virtual electromagnetic torque. This means that FVBs can be used to slow down or accelerate \acp{GFVSC} in the system.
Two control strategies based on~\cite{RAvilaM2022} will be analysed: 
\begin{itemize}
	\item Local fast voltage booster (FVB-L).
	\item Fast voltage booster using a wide-area control system (FVB-WACS). 
\end{itemize}

 Only a brief description of the FVBs are presented in this paper. The interested reader is referred to\cite{RAvilaM2022}. 
 
\subsection{Local fast voltage booster (FVB-L)}\label{sec:VSC_V_TS_FVB_L}
\begin{figure}[!htbp]
	\begin{center}
		\includegraphics[width=0.65\columnwidth]{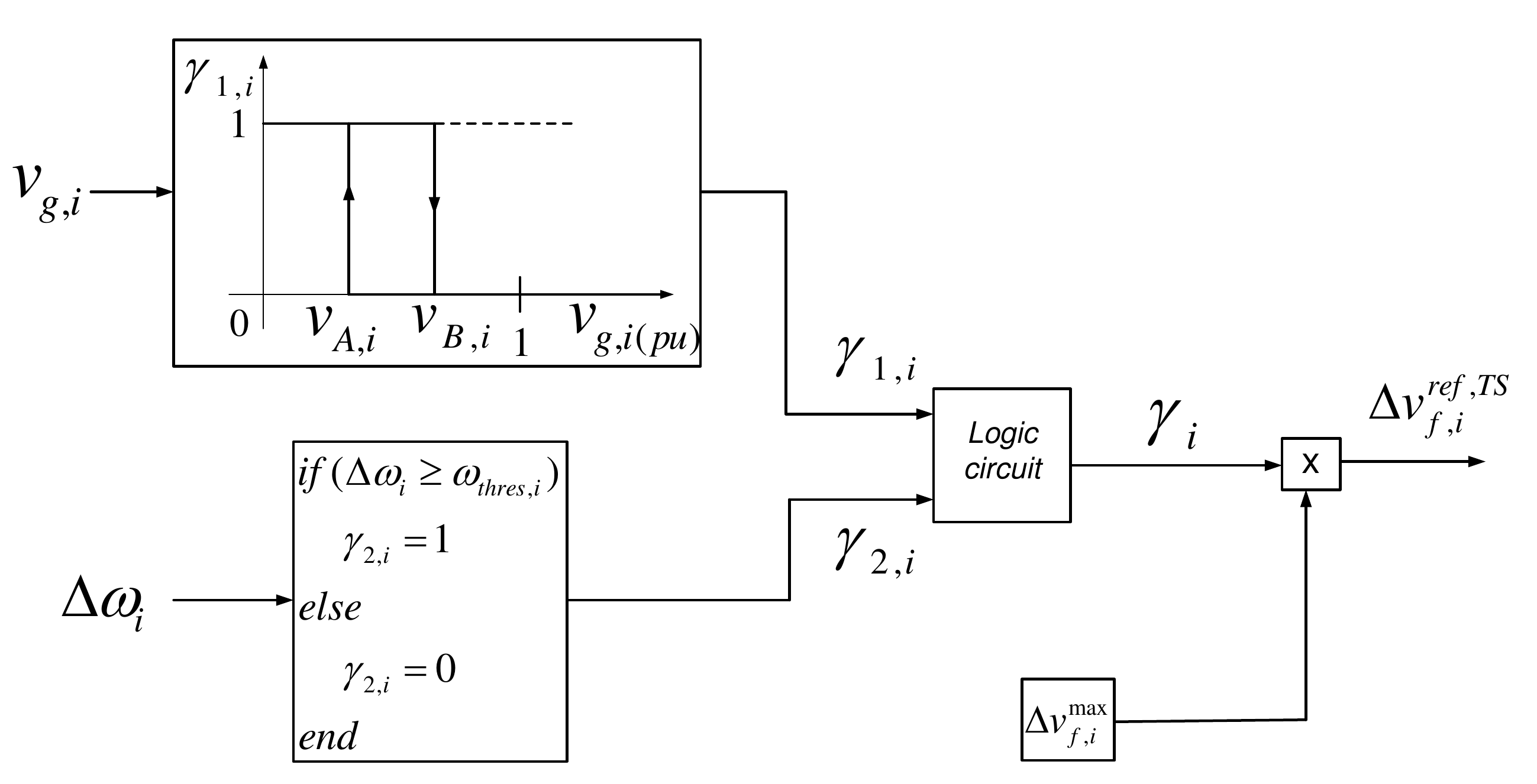}
		\caption{Strategy FVB-L. }
		\label{fig:VSC_GF_FVB_L}
		%\end{center}
		%\end{figure}
		%\begin{figure}[!htbp]
		%\begin{center}
		\includegraphics[width=0.65\columnwidth]{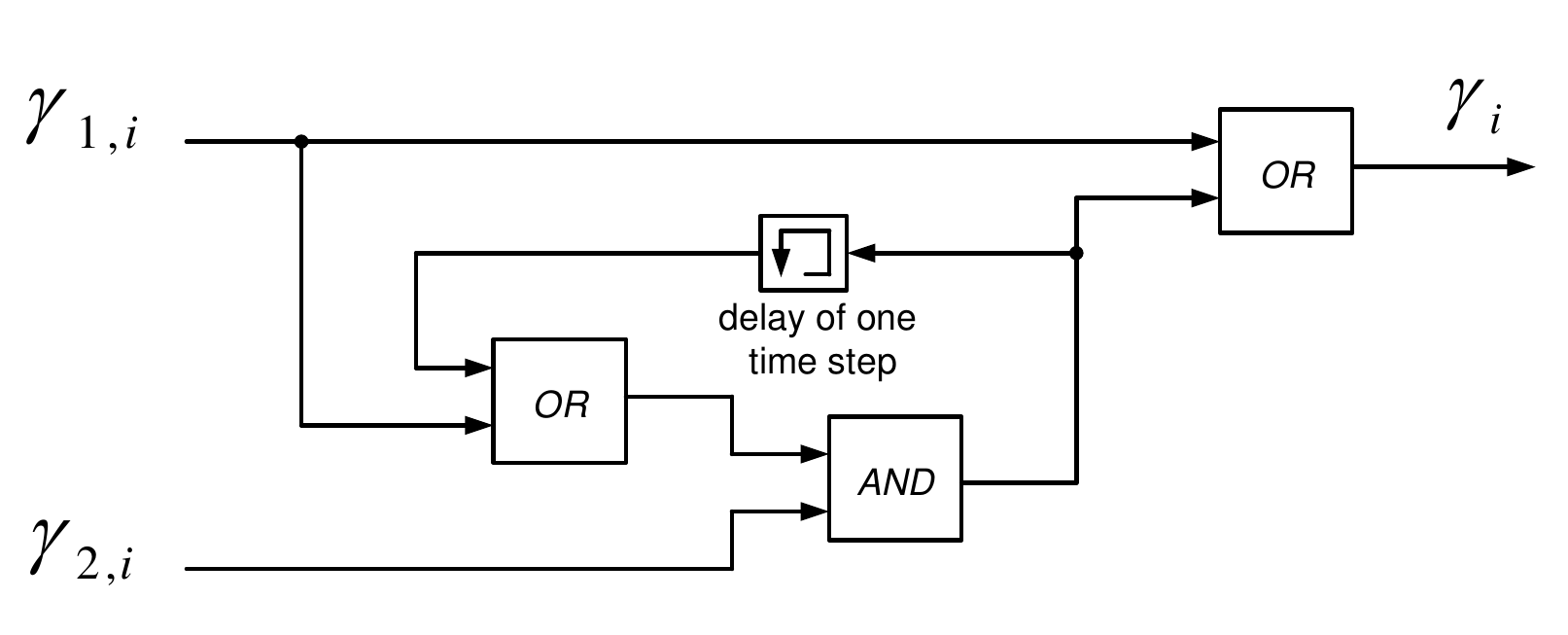}
		\setlength{\belowcaptionskip}{-10pt}
		\caption{Strategy FVB-L. Logic circuit for fault detection. }
		\label{fig:VSC_GF_FVB_L_logic}
	\end{center}
\end{figure}

\noindent This control strategy provides fast voltage support by using local measurements of the terminal voltage and the frequency deviation of each \ac{GFVSC}-$i$ ($v_{g,i}$ and $\Delta \omega_{i}$) as input signals, as shown in Figs.~\ref{fig:VSC_GF_FVB_L} and~\ref{fig:VSC_GF_FVB_L_logic}. Mainly, FVB-L is activated when the voltage at the connection point is lower than a certain threshold and it remains activated if the frequency difference is greater than a certain threshold ($v_{A,i}$, $v_{B,i}$ and $\omega_{thres,i}$). 
\vspace{-0.4cm}

 Transient stability is driven by the relative frequency in multi-converter systems with \acp{GFVSC}. This means that effective controls should slow down some \acp{GFVSC} but accelerate other ones. Since FVB-L uses local measurements only, the information of the rest of the system is not available. FVB-L strategy should be only activated in \acp{GFVSC} close to faults since these \acp{GFVSC} would accelerate faster, and they should be therefore slowed down. This is achieved by using a low value of parameter $v_{A,i}$ of Fig.~\ref{fig:VSC_GF_FVB_L} (e.g. $v_{A,i}=0.5$~pu)~\cite{RAvilaM2022}. 

%\FloatBarrier

\subsection{Fast voltage booster using a WACS (FVB-WACS)}\label{sec:VSC_V_TS_FVB_WACS}
\noindent This control strategy uses FVB support in each \ac{GFVSC} with a wide-area control system (WACS) to calculate the frequency of the COI, defined as: 
\begin{equation}\label{eq:w_COI}
	\omega_{COI}=\frac{1}{H_{tot}}\sum_{k=1}^{n} H_{VSC,k} \omega_{k} \mbox{ (pu)\space, with } H_{tot}= \sum_{k=1}^{n} H_{VSC,k}
\end{equation}
where $\omega_{k}$ is the frequency of each \ac{GFVSC} of the system and $H_{VSC,k}$ is its emulated inertia.

\begin{figure}[!htbp]
	\begin{center}
		\includegraphics[width=0.7\columnwidth]{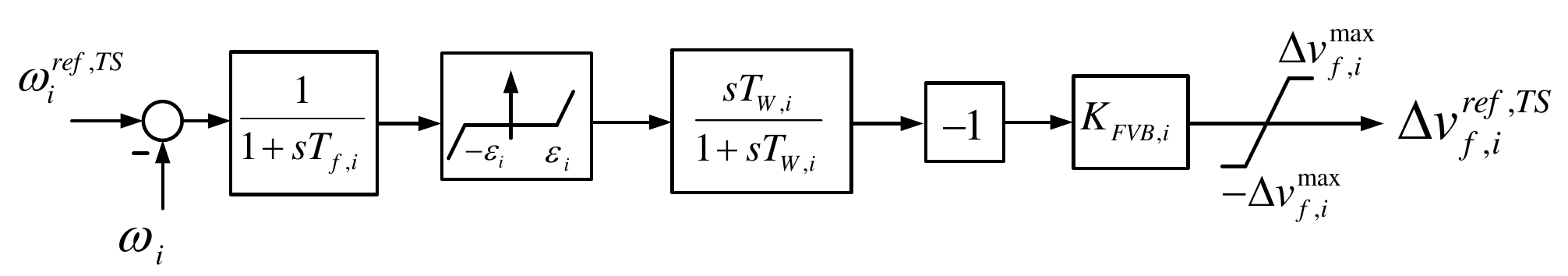}
		\caption{Strategy FVB-WACS. }
		\label{fig:VSC_GF_FVB_WACS}
	\end{center}
\end{figure}

The block diagram of the strategy FVB-WACS is depicted in Fig.~\ref{fig:VSC_GF_FVB_WACS}. The frequency set-point of each VSC-$i$ is calculated as the frequency of the COI from Eq.~(\ref{eq:w_COI}): $\omega_{i}^{ref,TS}=\omega_{COI}$.
The philosophy of the FVB-WACS control strategy is to increase (decrease) the voltage at the connection point of \ac{GFVSC}-$i$ if its frequency ($\omega_{i}$) is above (below) the frequency of the COI, in order to slow down (accelerate) the \ac{GFVSC}. The effect is to pull together the frequencies of the \acp{GFVSC} of the system. The strategy FVB-WACS is more powerful than strategy FVB-L, because it uses global measurements and, therefore, each \ac{GFVSC} has information of the rest of the system. Naturally, the implementation of FVB-WACS is more difficult, because a communication system is needed. 

\FloatBarrier

\section{Theoretical analysis}\label{sec:theo_analysis}
\noindent This section presents a theoretical analysis of the impact of FVBs and current limiters (CSA and HCL) in \acp{GFVSC} on transient stability by means of the well-known Equal Area Criterion (EAC)~\cite{Kundur1994a}. Analysis of active-power/angle ($P-\delta$) curves and EAC applied to transient stability assessment of \acp{GFVSC} has been used in previous work~\cite{Qoria_VSC_current_limit2020,Qoria_VSC_CCT2020,Guangya2021,XiongX_2021a,BlaabjergFTSAngle2022, SiW2023}.

\begin{figure}[!htbp]
	\begin{center}
		\centering
		\includegraphics[width=0.65\columnwidth]{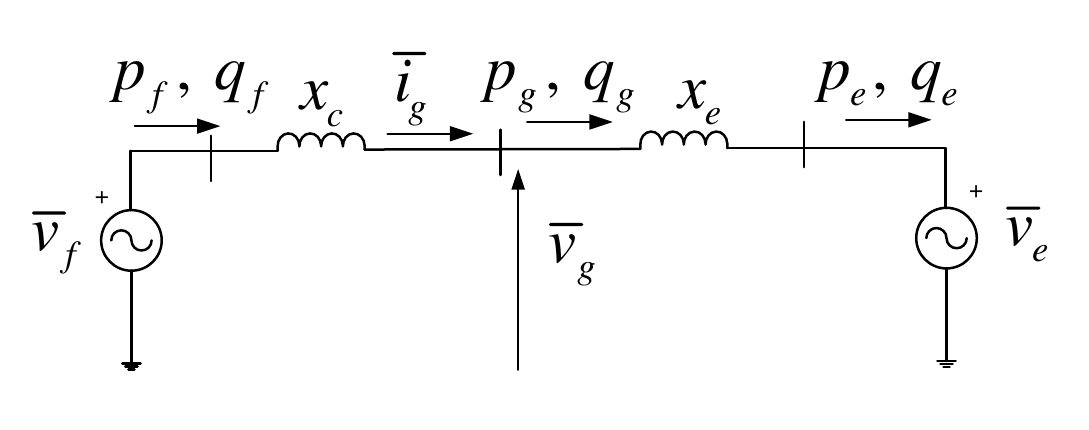}
		\caption{GFM-VSC connected to an infinite grid.}
		\label{fig:gfm_vsc_smib_sld}
	\end{center}
\end{figure}

A \ac{GFVSC} connected to an infinite grid is considered, as shown in Fig. \ref{fig:gfm_vsc_smib_sld}. The \ac{GFVSC} is represented as a voltage source at bus $f$ of Fig. \ref{fig:VSC_V_model} ($\bar{v}_f$), in order to analyse the impact of the FVBs. Some approximations are made to be able to carry out the theoretical analysis: series resistances are neglected, the shunt capacitor ($C_f$) is neglected (i.e., $\bar{i}_s \approx \bar{i}_{g}$) and the current and voltage controllers of the \ac{GFVSC} are not taken into account.

The infinite grid is the reference for the angles: 
$\bar{v}_e=v_e \angle 0$. The initial steady-state operating point of the \ac{GFVSC} is $\bar{s}^0=p_{g}^{0}+jq_{g}^{0}$ and $\bar{v}_{f}^{0}=v_{f}^{0} \angle \delta_{f}^{0}$. The active power injection of the \ac{GFVSC} can be written as:
\begin{equation}\label{eq:VSC_pg_EAC}
	p_{g} = p_{e}= \frac{v_{f} v_{e}}{x_{tot}} \sin \delta
\end{equation}
where $\delta = \delta_f$ is the angle of the voltage of the \ac{GFVSC} at bus $f$, $\bar{v}_f$, with respect to the infinite bus, and $x_{tot}=x_{c}+x_{e}$ is the total series reactance.

A three-phase-to-ground short circuit will be applied at bus $e$, for illustrative purposes. The \ac{GFVSC} has an initial angle of $\delta_0$. The fault is produced at time $t_{f}$ and it is cleared at $t_{cl}$. The angle of the \ac{GFVSC} at the fault clearing time will be called $\delta_{cl}$. During the fault, $v_e=0$, which implies $p_g=0$. Hence, according to the VSM law of the \ac{GFVSC} (Fig. \ref{fig:VSC_V_VSM}), during the fault the frequency of the \ac{GFVSC} accelerates according to:
\begin{eqnarray}\label{eq_VSC_SM_EAC}
	2 H_{VSC}\frac{d \Delta \omega}{dt} + D_{VSC} \Delta \omega & = & p_{g}^{0} - \cancelto{0}{p_{g,i}}  
\end{eqnarray}

\subsection{Impact of current limiters on transient stability}\label{sec:theo_analysis_GFM_CL}
The base case is considered, where the AC voltage of the \ac{GFVSC} is assumed constant: $v_f=v_f^{0}$. 
\subsubsection{Base case - no CL}\label{sec:theo_GFM_base_no_CL}
\noindent First of all, and for comparison purposes, it is assumed that the \ac{GFVSC} does not have current limiter (CL) and its active-power injection is given by: 
\begin{equation}\label{eq:GFM_pg_case1}
	p_{g} = \frac{v_{f}^{0} v_{e}}{x_{tot}} \sin \delta
\end{equation}
The the $P-\delta$ curve of the \ac{GFVSC} is shown in Fig. \ref{fig:EAC_case1}-(a). The maximum active-power injection is $p_{g}^{max1}=v_{f}^{0} v_{e}/x_{tot}$. The \ac{GFVSC} is initially operating with an angle $\delta_0$ and active-power injection $p_g^0$. The fault is cleared when the angle reaches $\delta_{cl}$. The accelerating area is called $A_1$, and it is shown in shaded red. To maintain synchronism, the decelerating area $A_2$ (shaded blue) needs to be equal to the accelerating area $A_1$, which means that the decelerating energy is equal to the energy increase during the fault. During the re-synchronisation process a maximum transient excursion angle is reached ($\delta_{mte}$). 

The critical clearing angle ($\delta_{cl}^{crit}$) is the maximum angle that can be achieved before clearing the fault without loss of synchronism, which is directly related to the critical clearing time of the fault. The EAC criterion for the critical clearing angle is illustrated in Fig. \ref{fig:EAC_case1}-(b), where $\delta_{mte}^{crit}=\pi-\delta_0$.

\subsubsection{Base case - CSA}\label{sec:theo_GFM_base_CSA}
\noindent The \ac{GFVSC} has now a current limiter with CSA algorithm whith equal priority to $d-$ and $q-$axis currents (Section \ref{sec:VSC_V_current_lim_CSA}). The current limit is reached during the fault and the instants after the fault clearing. When the fault is cleared, the angle reaches the value $\delta_{cl}$. Angle $\delta_a$ represents the angle at which the current limit ($i_s^{max}$) is reached. From that angle, the $P-\delta$ curve is no longer a sinusoidal function, because the \ac{GFVSC} has the current injection saturated. The active-power injection of the \ac{GFVSC} is given by~\cite{Guangya2021}: 
\begin{equation}\label{eq:GFM_pg_case2}
\begin{split}
    p_{g} = \left\lbrace\begin{array}{ccc} 
        \frac{v_{f}^{0} v_{e}}{x_{\text{tot}}} \sin \delta, &\text{si}& \delta < \delta_a (i_s < i_s^{\max}) \\\\
        \frac{v_{f}^{0} v_{e} \sin \delta}{\sqrt{(v_{f}^{0})^2 + v_{e}^2-2v_{f}^{0} v_{e} \cos \delta}} \cdot i_{s}^{\max}, &\text{si}& \delta \geq \delta_a (i_s = i_s^{\max})
    \end{array}\right.
\end{split}
\end{equation}

Fig. \ref{fig:EAC_case2}-(a) shows The the $P-\delta$ curve of the \ac{GFVSC} with CSA. Clearly, the $P-\delta$ curve is reduced dramatically and a higher maximum transient excursion angle $\delta_{mte}$ is required to achieve $A_1=A_2$ and prevent loss of synchronism of the \ac{GFVSC}, in comparison with the case of no CL. The maximum active-power injection is given by $p_{g}^{max2}=v_{f}^{0} v_{e} \sin(\delta_c)/x_{tot}$, which is much lower than $p_{g}^{max1}$. This results in a much lower critical clearing angle $\delta_{cl}^{crit}$ (\ref{fig:EAC_case2}-(a)). Hence, transient stability margin of the \ac{GFVSC} with CSA is much lower, in comparison with the case without CL (see~\cite{Qoria_VSC_current_limit2020,Qoria_VSC_CCT2020,Guangya2021}).

\subsubsection{Base case - HCL}\label{sec:theo_GFM_base_HCL}
\noindent The \ac{GFVSC} has now a current limiter with HCL algorithm (Section \ref{sec:VSC_V_current_lim_hybrid_CL}). As explained before, the performance of HCL current limiter in terms of transient stability is governed by VI current limiter and CSA algorithm just ensures that the current injection of the \ac{GFVSC} is not greater that the current limit during specific moments of the transient. Hence, for this analysis, it it assumed that the \ac{GFVSC} is in VI current-limitation mode when it saturates (Section \ref{sec:VSC_V_current_lim_VI}). 

To simplify the analysis, it assumed that the virtual impedance included when the current threshold is reached is constant and the virtual resistance is neglected: $\bar{z}_{VI}=j x_{VI}$, with $x_{VI}=k_{p_{r_{VI}}} \sigma_{x/r} \Delta i_{s}^{max}$ and $\Delta i_{s}^{max}=i_{s} - i_{thres}^{VI}$.

When HCL limiter is activated, the virtual voltage of the \ac{GFVSC} is given by $\bar{v}_{f}^{'} = \bar{v}_{f} + \bar{z}_{VI} \bar{i}_{s}$ with $\bar{v}_{f}^{'}= v_{f}^{'} \angle \delta_{f}^{'}$. Let us define the angle $\delta$ as the angle of the virtual voltage of the \ac{GFVSC} with respect to the infinite grid: $\delta = \delta_{f}^{'}$. Notice, that when the HCL is not activated (virtual impedance equal to zero: $\bar{z}_{VI}=0$) leads to $\bar{v}_{f}^{'}=\bar{v}_{f}$ and, therefore, the angle $\delta$ is the angle of the AC voltage at bus $f$, as in previous subsections. For mathematical purposes, the virtual voltage is estimated as:
\begin{equation}\label{eq:GFM_HCL_vfp_0}
	\bar{v}_{f}^{0'} = \bar{v}_{f}^{0'} + \bar{z}_{VI} \bar{i}_{s}^{0'}
\end{equation}
where the current used for the calculation, $\bar{i}_{s}^{0'}$, is an approximation with maximum magnitude $i_{s}^{max}$ and equal phase than the current at the initial operating point.

Then, the active-power injection of the \ac{GFVSC} is given by:  
\begin{eqnarray}\label{eq:GFM_pg_case3}
    p_{g} = \left\lbrace\begin{array}{ccc} 
        \frac{v_{f}^{0} v_{e}}{x_{tot}} \sin \delta, &if& \delta \leq \delta_b (i_s \leq i_{thres}^{VI})
        \\\\
        \frac{v_{f}^{'} v_{e}}{x_{tot}+x_{VI}} \sin \delta, &if& \delta > \delta_b (i_s > i_{thres}^{VI}) 
        \end{array}\right.
\end{eqnarray}
where $\delta_b$ is the angle at which the VI current threshold is reached and the maximum value of the curve is $p_{g}^{max3}=v_{f}^{0'} v_{e}/(x_{tot}+x_{VI})$.

\begin{figure}[!htbp]
	\begin{center}
		\centering
		\includegraphics[width=0.8\columnwidth]{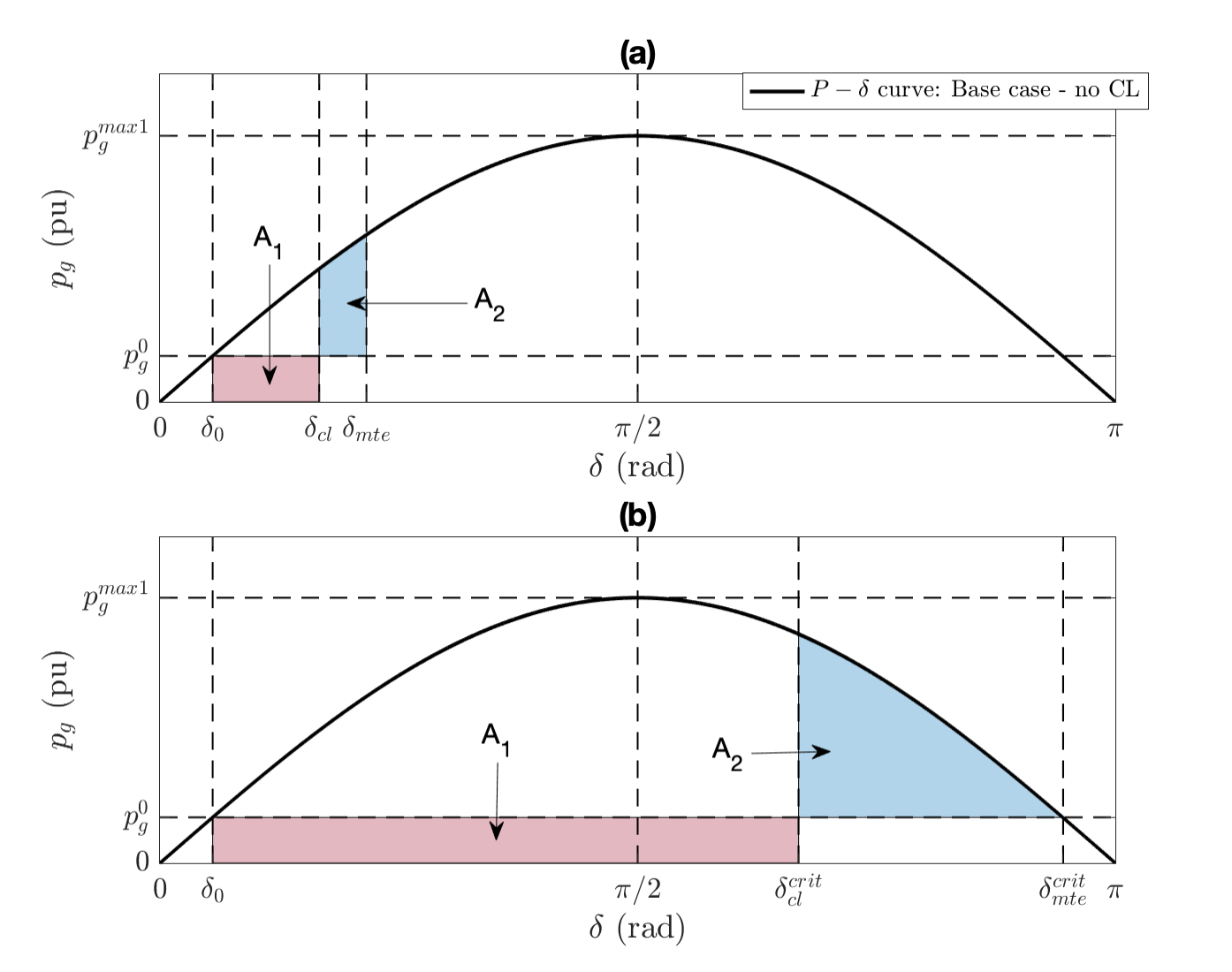}
		\caption{Base case - no CL: $P-\delta$ curve and EAC.}
		\label{fig:EAC_case1}
	\end{center}
% \end{figure}
% \begin{figure}[!htbp]
	\begin{center}
		\centering
		\includegraphics[width=0.8\columnwidth]{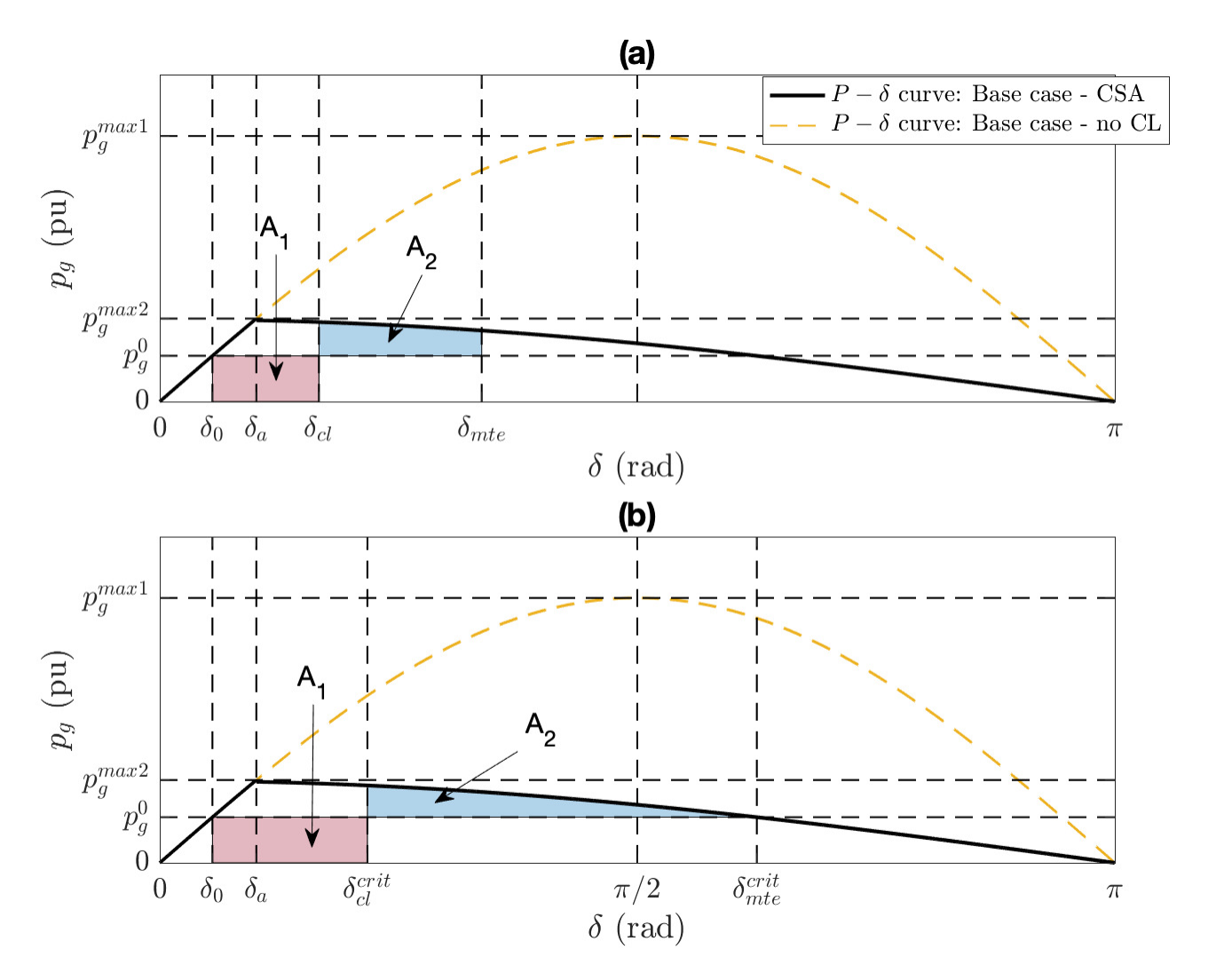}
		\caption{Base case - CSA: $P-\delta$ curve and EAC.}
		\label{fig:EAC_case2}
	\end{center}
\end{figure}

\begin{figure}[!htbp]
	\begin{center}
		\centering
		\includegraphics[width=0.8\columnwidth]{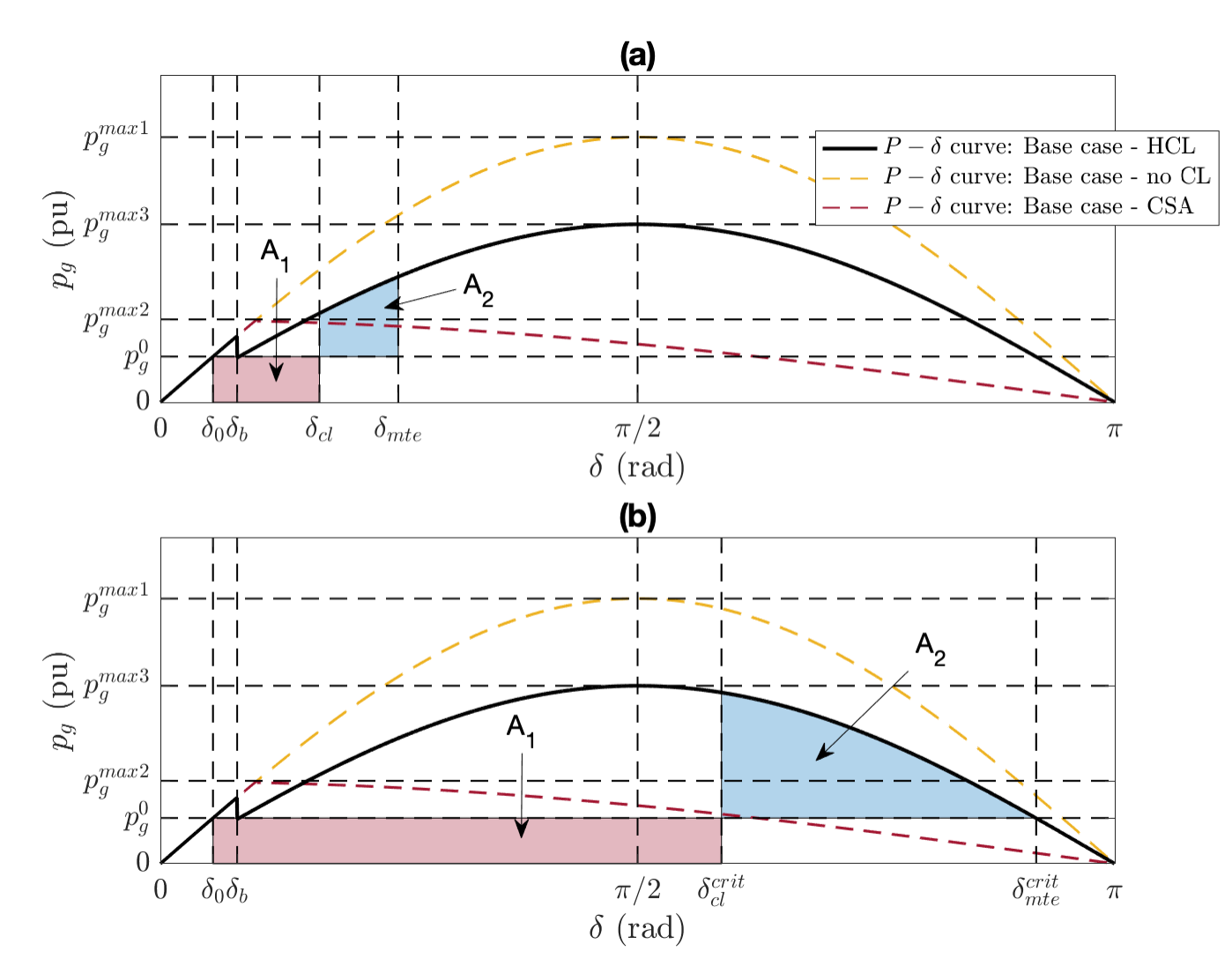}
		\caption{Base case - HCL: $P-\delta$ curve and EAC.}
		\label{fig:EAC_case3}
	\end{center}
% \end{figure}
% \begin{figure}[!htbp]
	\begin{center}
		\centering
		\includegraphics[width=0.8\columnwidth]{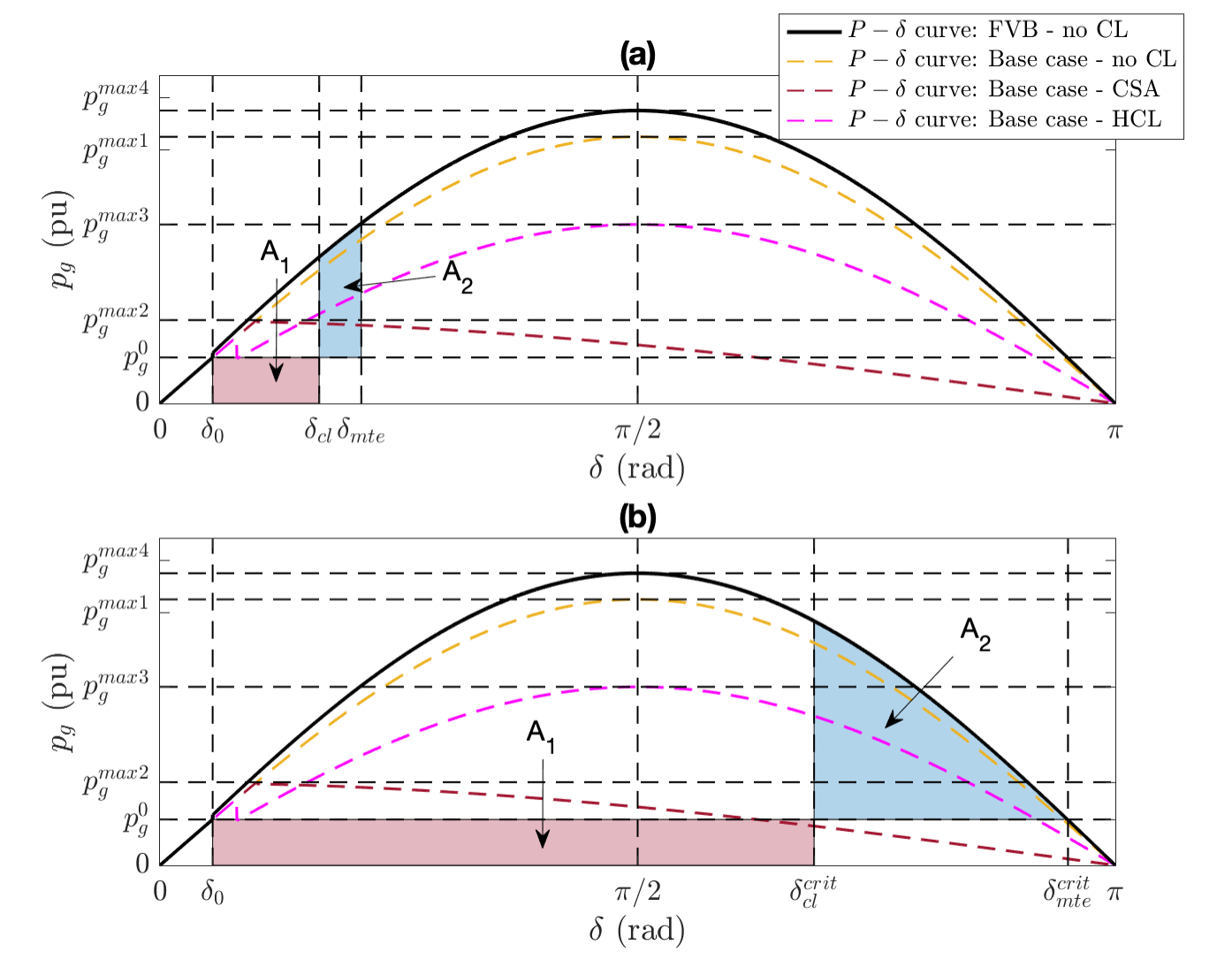}
		\caption{FVB - no CL: $P-\delta$ curve and EAC.}
		\label{fig:EAC_case4}
	\end{center}
\end{figure}

Fig. \ref{fig:EAC_case3}-(a) shows The the $P-\delta$ curve of the \ac{GFVSC} with HCL. Notice that there is a discontinuity of the active power function at $\delta_b$, due to the introduction of the virtual impedance of the HCL. Clearly, the $P-\delta$ curve with HCL is larger than the one obtained in the case of \ac{GFVSC} with CSA (Fig. \ref{fig:EAC_case2}), and the \ac{GFVSC} has more margin to maintain synchronism by achieving the EAC: $A_2=A_1$. As a consequence, the critical clearing angle ($\delta_{cl}^{crit}$) obtained with HCL (Fig. \ref{fig:EAC_case3}-(b)) is much greater than the one obtained with CSA (Fig. \ref{fig:EAC_case2}-(b)). Naturally, the $P-\delta$ curve with HCL (Fig. \ref{fig:EAC_case3}) is smaller than the one obtained in the case of no current limitation (Fig. \ref{fig:EAC_case1}), having the former a lower critical clearing angle $\delta_{cl}^{crit}$.

Further details of theoretical analysis of transient stability of \acp{GFVSC} using VI current limiters can be found in~\cite{Qoria_VSC_current_limit2020,Qoria_VSC_CCT2020}.

\begin{figure}[!htbp]
	\begin{center}
		\centering
		\includegraphics[width=0.8\columnwidth]{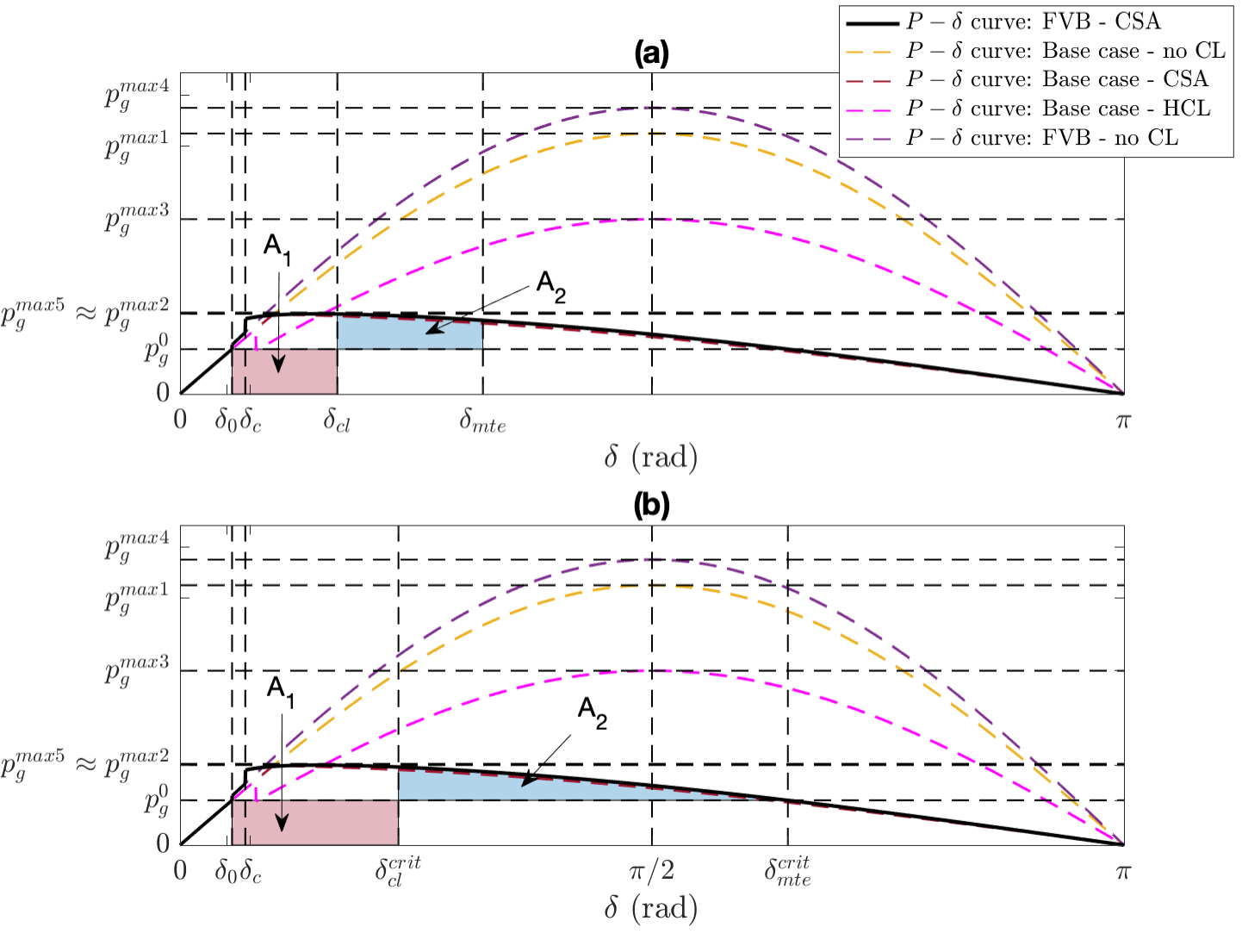}
		\caption{FVB - CSA: $P-\delta$ curve and EAC.}
		\label{fig:EAC_case5}
	\end{center}
%\end{figure}
%\begin{figure}[!htbp]
	\begin{center}
		\centering
		\includegraphics[width=0.8\columnwidth]{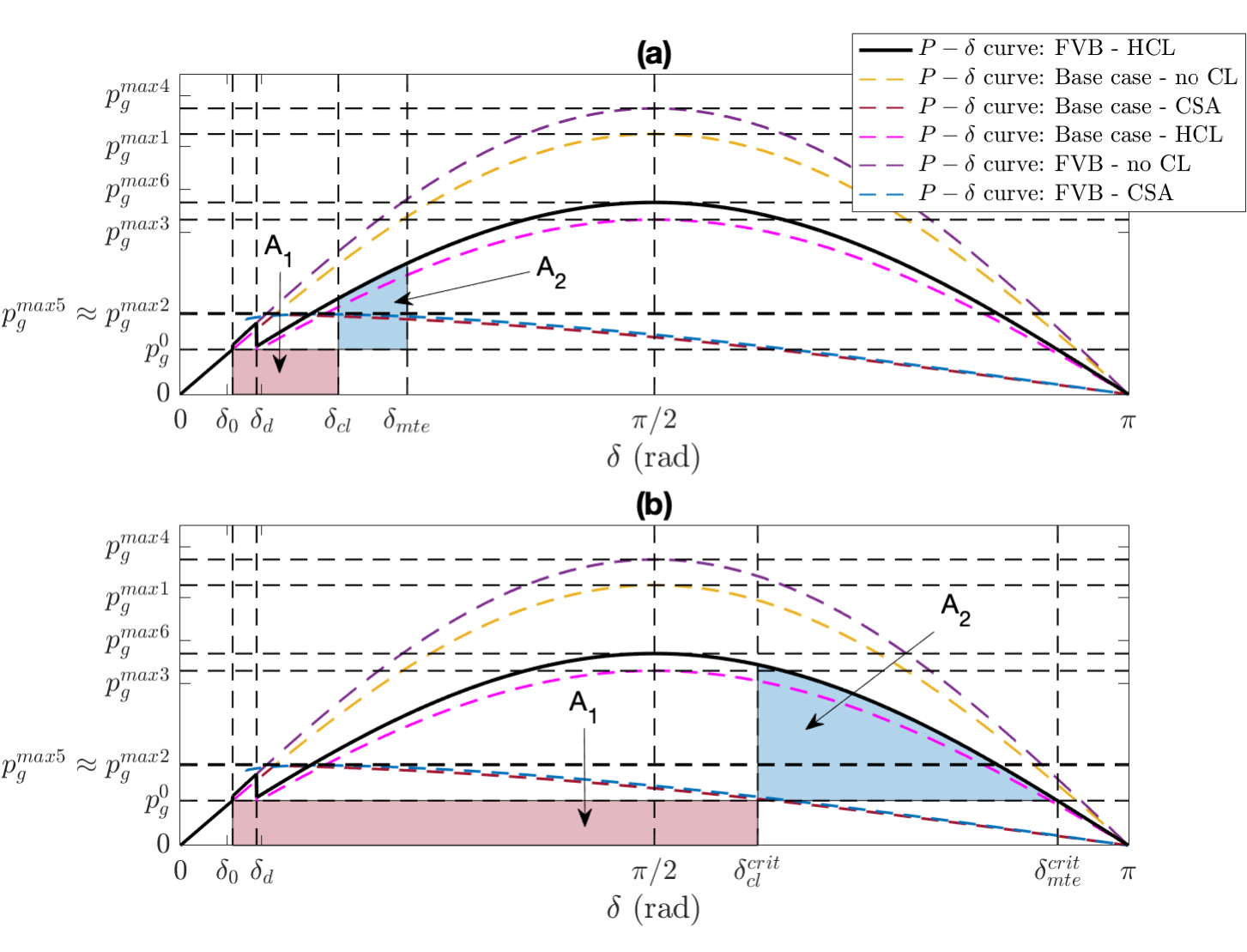}
		\caption{FVB - HCL: $P-\delta$ curve and EAC.}
		\label{fig:EAC_case6}
	\end{center}
\end{figure}

\FloatBarrier
\subsection{Impact of FVBs on transient stability}\label{sec:theo_analysis_GFM_FVB}
\noindent The \ac{GFVSC} is now equipped with a FVB as the ones described in Section \ref{sec:VSC_V_TS}. Hence, the AC voltage of the \ac{GFVSC} is given by: $v_f=v_f^{0}+\Delta v_f^{TS}$
\begin{equation}\label{eq:GFM_EAC_FVB_vf}
	v_f=v_f^{0}+\Delta v_f^{TS}
\end{equation}
where the term $\Delta v_f^{TS}$ is the voltage increment produced by the FVB.

Again, certain assumptions are made, in order to carry out the theoretical analysis. A positive voltage increment produced by the FVB ($\Delta v_f^{TS}>0$) when the fault is detected is assumed. This is true for FVB-L and FVB-WACS (see Section \ref{sec:VSC_V_TS}) in case of faults close to the AC connection point of the \ac{GFVSC}, because the fault will produce that the frequency of the \ac{GFVSC} is greater than the frequency of the COI. The second assumption is a constant voltage increment produced by the FVB ($\Delta v_f^{TS}=\Delta v_f^{TS,cons}$). This is true for FVB-L, but not for FVB-WACS, because the latter modulates the supplementary voltage set-point according to frequency measurements in the system. Nevertheless, the impact of generic FVBs with constant supplementary voltage set-point is useful to illustrate their impact on transient stability, which is the purpose of the analysis; and the general conclusions are valid for both FVB-L and FVB-WACS. The constant voltage used for the $P-\delta$ curves when the FVB is applied is called $v_f^{1}=v_f^{0}+\Delta v_f^{TS}$

\subsubsection{FVB - no CL}\label{sec:theo_GFM_fvb_no_CL}
\noindent The \ac{GFVSC} does not have current limiter (CL) and a FVB is used. Then, its active-power injection is given by: 
\begin{equation}\label{eq:GFM_pg_case1}
	p_{g} = \frac{v_{f}^{1} v_{e}}{x_{tot}} \sin \delta
\end{equation}
where $v_f^{1}=v_f^{0}+\Delta v_f^{TS}$. 

Fig. \ref{fig:EAC_case4} shows The the $P-\delta$ curve and the EAC. The maximum active-power injection is $p_{g}^{max4}=v_{f}^{0} v_{e}/x_{tot}>p_{g}^{max1}$. With FVB, the critical clearing angle $\delta_{cl}^{crit}$ is greater than the one obtained in the base case (\ref{sec:theo_analysis_GFM_CL}), because the AC voltage of the \ac{GFVSC} is higher $v_f^{1}>v_f^{0}$ and the $P-\delta$ curve is higher.

\subsubsection{FVB - CSA}\label{sec:theo_GFM_fvb_CSA}
\noindent The \ac{GFVSC} has now a current limiter with CSA algorithm with equal priority to $d-$ and $q-$axis currents (Section \ref{sec:VSC_V_current_lim_CSA}) and a FVB. Angle $\delta_c$ represents the angle at which the current limit ($i_s^{max}$) is reached. From that angle, the $P-\delta$ curve is no longer a sinusoidal function, because the \ac{GFVSC} has the current injection saturated. The active-power injection of the \ac{GFVSC} is given by the same expression as in (\ref{eq:GFM_pg_case2}), with the difference that now the \ac{GFVSC} has a higher AC voltage due to the FVB ($v_f^{1}=v_f^{0}+\Delta v_f^{TS}$): 
\begin{equation}\label{eq:GFM_pg_case5}
\begin{split}
    p_{g} = \left\lbrace\begin{array}{ccc} 
        \frac{v_{f}^{1} v_{e}}{x_{tot}} \sin \delta, &if& \delta < \delta_c (i_s < i_s^{max})
        \\\\
        \frac{v_{f}^{1} v_{e} \sin \delta}{\sqrt{(v_{f}^{1})^2 + v_{e}^2-2v_{f}^{1} v_{e} \cos \delta}} \cdot i_{s}^{max}, &if& \delta \ge \delta_c (i_s = i_s^{max}) 
        \end{array}\right.
\end{split}
\end{equation}

Fig. \ref{fig:EAC_case5} shows The the $P-\delta$ curve and the EAC. The $P-\delta$ curve is reduced dramatically and the FVB is not capable of increasing significatly the $P-\delta$. Although the $P-\delta$ curve obtained with CSA and FVB is higher than the one obtained in the base case with CSA (Fig. \ref{fig:EAC_case2}), both are very similar. In fact, their maximum values of the curve are almost equal $p_{g}^{5}\approx p_{g}^{5}$ and the critical clearing angles $\delta_{cl}^{crit}$ are similar. Although it will be shown in a numerical example that using FVB the critical clearing angle increases slightly with FVB (Section \ref{sec:theo_analysis_GFM_results}).

\subsubsection{FVB - HCL}\label{sec:theo_GFM_fvb_HCL}
\noindent The \ac{GFVSC} has now a current limiter with HCL algorithm (Section \ref{sec:VSC_V_current_lim_hybrid_CL}) and FVB. The virtual impedance is activated during the fault and in the instants after the fault clearing, when the current excess the threshold (see Section \ref{sec:VSC_V_current_lim_VI}). The virtual voltage used for the $P-\delta$ curve is estimated as:
\begin{equation}\label{eq:GFM_HCL_vfp_0}
	\bar{v}_{f}^{'1} = \bar{v}_{f}^{1} + \bar{z}_{VI} \bar{i}_{s}^{0'}
\end{equation}
where the AC voltage of the \ac{GFVSC} is $v_f^{1}=v_f^{0}+\Delta v_f^{TS}$.

The active-power injection of the \ac{GFVSC} is given by the same expression as in (\ref{eq:GFM_pg_case3}), with the difference that now the \ac{GFVSC} has a higher AC voltage due to the FVB ($v_f^{1}=v_f^{0}+\Delta v_f^{TS}$) and a higher virtual voltage ($v_f^{1'}$):
\begin{eqnarray}\label{eq:GFM_pg_case6}
    p_{g} = \left\lbrace\begin{array}{ccc} 
        \frac{v_{f}^{1} v_{e}}{x_{tot}} \sin \delta, &if& \delta \leq \delta_d (i_s \leq i_{thres}^{VI})
        \\
        \frac{v_{f}^{1'} v_{e}}{x_{tot}+x_{VI}} \sin \delta, &if& \delta > \delta_d (i_s > i_{thres}^{VI}) 
        \end{array}\right.
\end{eqnarray}
where $\delta_d$ is the angle at which the VI current threshold is reached and the maximum value of the curve is $p_{g}^{max6}=v_{f}^{0'} v_{e}/(x_{tot}+x_{VI})>p_{g}^{max3}$.

Fig. \ref{fig:EAC_case3} shows The the $P-\delta$ curve and the EAC of the \ac{GFVSC} with HCL and FVB. The critical clearing angle ($\delta_{cl}^{crit}$) obtained with FVB and HCL is greater than the one obtained in the base case with HCL.

\subsection{Numerical example}\label{sec:theo_analysis_GFM_results}
The system of Fig. \ref{fig:gfm_vsc_smib_sld} is considered (a \ac{GFVSC} connected to an infinite grid). The rating of the \ac{GFVSC} is 900 MVA and its data are the ones of VSC1 of Table \ref{tab:VSC_parameters} of the Appendix (one of the converters to be used in the results' section of this paper). The infinite grid has voltage $\bar{v}_e=1 \angle 0^o$ and series reactance $\bar{z}_e=j x_{e}=j 0.1$ pu. The operating point of the \ac{GFVSC} is $p_g^0=0.7$ pu, $q_g^0=0.0490$ pu, $v_f^{0}=1.0152$ pu and $\delta_0=\delta_f^0=9.93^o$. For the CSA current limiter, a maximum current of $i_s^{max}=1.25$ pu is assumed. For the HCL current limiter, a current threshold of $i_{VI,i}^{max}=1.0$~pu, rest of parameters as in the Appendix, the resistance is neglected and a constant impedance is used for the EAC analysis, as explained in Section \ref{sec:theo_GFM_base_HCL}. This leads to a virtual impedance of $\bar{z}_{VI}=j x_{VI}=0.1225$ pu. The voltage increment of the FVB is $\Delta v_f^{TS}=0.1$ pu. All the parameters are in pu with respect to the converter rating.

Table \ref{tab:EAC_results} summarises the main results of the EAC analysis: the critical clearing angle, which is used to quantify transient-stability margin, and the maximum value of the $P-\delta$ curve, for each case. In the base case without CL, the critical clearing angle is $\delta_{cl}^{crit}=120.32^o$, and it decreases to $39.13^o$ with CSA. With HCL, $\delta_{cl}^{crit}=105.83^o$, significantly increasing transient stability margin in comparison with the base case with CSA.

With FVB and without CL, the critical clearing angle is $\delta_{cl}^{crit}=123.24^o$, which is slightly higher than the one obtained in the base case. With FVB and CSA, the critical clearing angle decreases to $41.65^o$, which is slightly higher than the one obtained in the base case with CSA. With FVB and HCL, $\delta_{cl}^{crit}=109.61^o$, which is higher than the one obatined with FVB and HCL. 

The analysis shows that HCL current limiters and FVBs increase transient stability margin, and both approaches are compatible. The improvements are higher with the HCLs. It should be mention that time-domain simulation have shown higher improvements with FVBs in comparison with the base case \cite{RAvilaM2022}. This is due to the fact that the main improvements of FVBs are due to the control actions after the current limitation. In the EAC analysis, the \ac{GFVSC} reaches its current limit during the complete post-fault re-synchronisation process, because its controllers are neglected. This does not occur in time-domain simulations, where greater improvements of FVBs are expected, as will be analysed in Section \ref{sec:Results2_VSC_V_Kundur} of this paper

\begin{table}[!htbp]
	\begin{center}
    
		\caption{EAC. Results.}
			\begin{tabular}{|l|c|c|}
				\hline
				Case & $\delta_{cl}^{crit}$   & $p_g^{max}$     \\ \hline
				Base case - no CL   & $120.32^o$ &     $p_{g}^{max1}=4.06$ pu \\                  
				Base case - CSA   & $39.13^o$ &     $p_{g}^{max2}=1.27$ pu \\                    
    			Base case - HCL   & $105.83^o$&     $p_{g}^{max3}=2.73$ pu \\  
       		FVB - no CL   & $123.24^o$ &     $p_{g}^{max4}=4.46$ pu \\                  
				FVB - CSA   & $41.65^o$&     $p_{g}^{max5}=1.25$ pu \\                          
    			FVB - HCL   & $109.61^o$&     $p_{g}^{max6}=2.99$ pu \\  
				\hline
			\end{tabular}
		\label{tab:EAC_results}
\end{center}
\end{table}

\FloatBarrier
\section{Results}\label{sec:Results2_VSC_V_Kundur}
\begin{figure}[!htbp]
	%\begin{center}
	\centering
	\includegraphics[width=1.0\columnwidth]{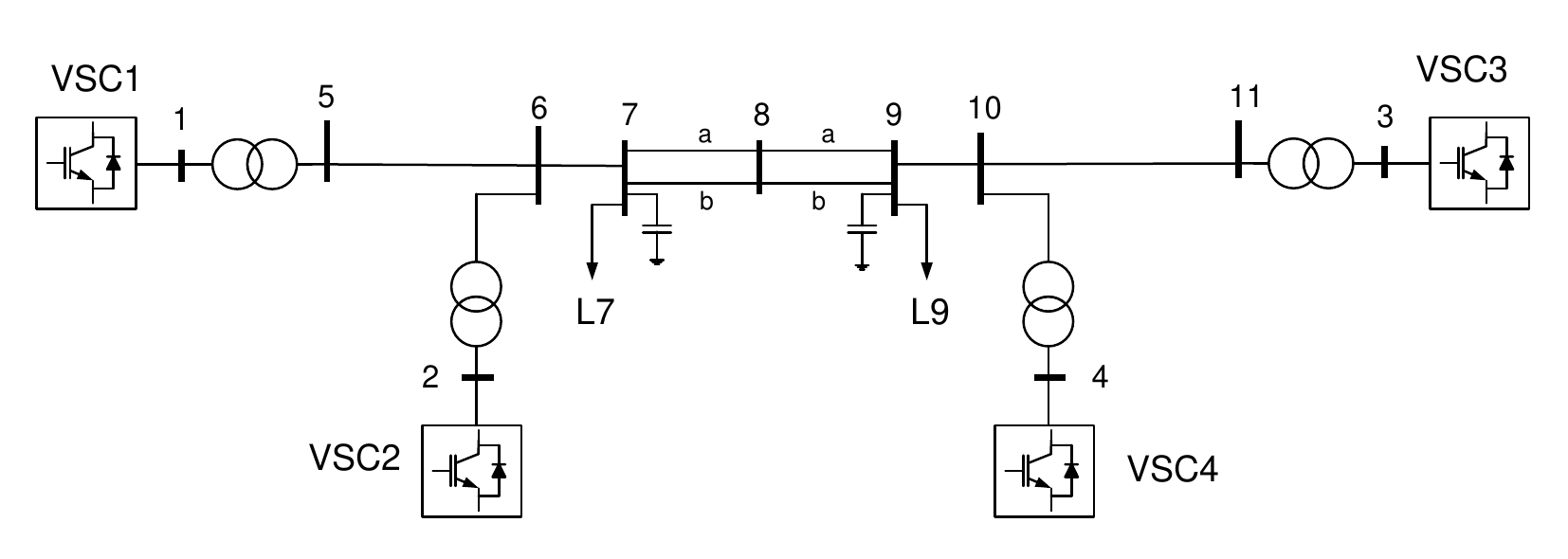}
	\caption{Kundur's two-area test system with 100~\% \ac{GFVSC}-based generation. }
	\label{fig:Kundur_two_area_VSC_V}
	%\end{center}
\end{figure}

\noindent Simulations were carried out to analyse the impact of the simultaneous use of FVBs and HCLs on Kundur’s two-area test system~\cite{Kundur1994a} with 100\% grid-forming VSC-based generation (see Fig.~\ref{fig:Kundur_two_area_VSC_V}). Synchronous machines of the original system were replaced by \ac{GFVSC}-based generators using VSM control, with the same nominal apparent power as the generators of the original system (900 MVA). The Appendix provides the system data. The VSC\_Lib tool was used: an open-source tool based on Matlab + Simulink + SimPowerSystems developed by L2EP-LILLE~\cite{L2EP_VSC_GF_2020,MIGRATE_WP3_2018a,Qoria2019a}, which uses average electromagnetic-type models.

The following cases will be analysed and compared:
\begin{itemize}
	\item Base case: current limiter only (CSA or HCL).
    \item FVB-L (local) and current limiter (CSA or HCL). 
    \item FVB-WACS (global) and current limiter (CSA or HCL). 
\end{itemize}

Parameters of of the FVBs and current limiters are provided in the Appendix. Note that VI-CL current limitation strategy is not analysed, since HCL presents better behaviour in terms of current limitation as shown in~\cite{Qoria_VSC_current_limit2020}.

\FloatBarrier
\subsection{Short-circuit simulation}\label{sec:Results2_VSC_V_Kundur_sim1}
\noindent A three-phase-to-ground short circuit is applied to line 7-8$a$ (close to bus 7) of the system in Fig.~\ref{fig:Kundur_two_area_VSC_V}, and the fault is cleared by disconnecting the circuit 140~ms later. 

\begin{figure}[!htbp]
	\begin{center}
		\centering
		\includegraphics[width=0.65\columnwidth]{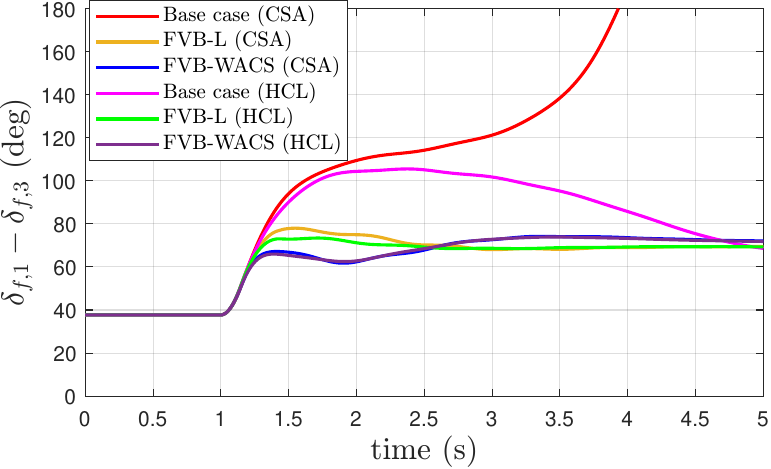}
		\caption{Fault I cleared after 140~ms. Angle difference of the VSCs. }
		\label{fig:Kundur_sim2_angles}
	\end{center}
\end{figure}

\begin{figure}[!htbp]
	\begin{center}
		\centering
		\vspace{-0.1cm}
		\includegraphics[width=0.65\columnwidth]{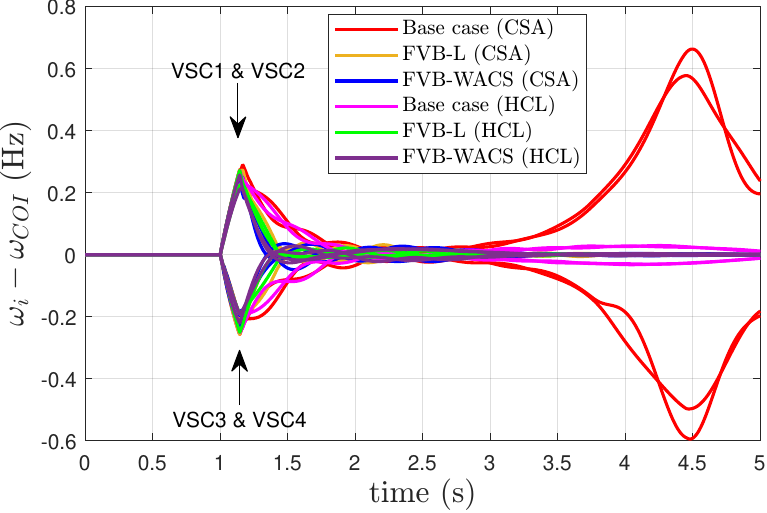}
		\setlength{\belowcaptionskip}{-10pt}
		\caption{Fault I cleared after 140~ms. Frequency deviations of the VSCs with respect to the frequency of the COI.}
		\label{fig:Kundur_sim2_freq_COI}
		\end{center}
\end{figure}

Fig.~\ref{fig:Kundur_sim2_angles} shows the angle difference between \ac{GFVSC}-1 and \ac{GFVSC}-3 for the six cases analysed, while Fig.~\ref{fig:Kundur_sim2_freq_COI} shows the frequency deviations of the \acp{GFVSC} with respect to the frequency of the COI. In the base case with CSA and no FVBs, \acp{GFVSC} lose synchronism. The system maintains synchronism for the rest of the cases (see Fig.~\ref{fig:Kundur_sim2_angles}). 
Results show that strategies FVB-L and FVB-WACS improve transient stability when using CSA and HCL current limiters.

\begin{figure}[!htbp]
		\begin{center}
		\centering
		\vspace{-0.1cm}
		\includegraphics[width=1.05\columnwidth]{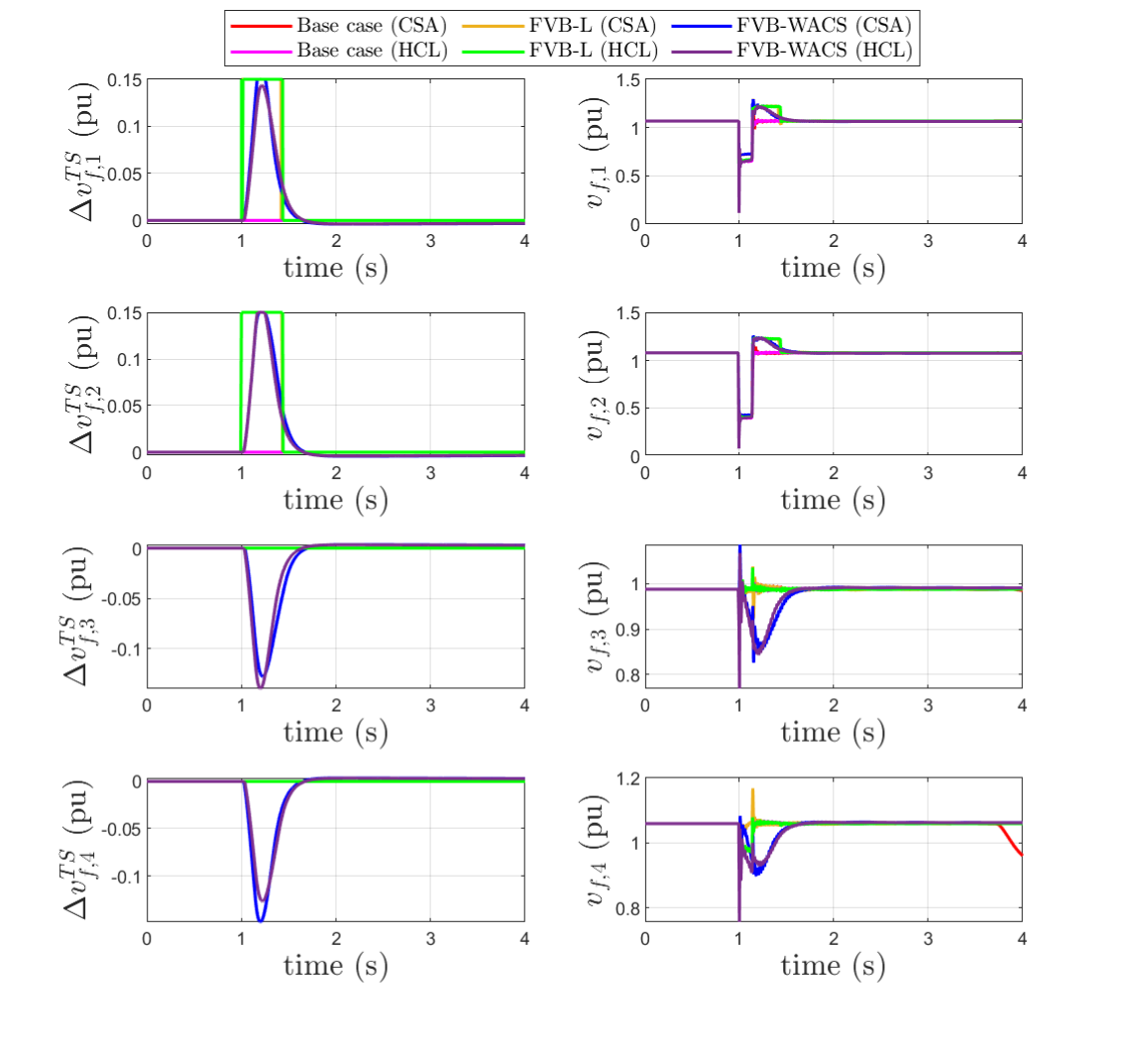}
		\vspace{-0.5cm}
		\caption{Fault I cleared after 140~ms. (left) Supplementary voltage set-points of the VSCs and (right) voltages of the VSCs. }
		\label{fig:Kundur_sim2_bVf_TS}
		\end{center}
\end{figure}

Fig.~\ref{fig:Kundur_sim2_bVf_TS} shows the supplementary voltage set-point (provided by FVBs) and the output voltages of each \ac{GFVSC}. The FVB-L strategy provides a positive supplementary voltage set-point in all converters that detect the fault. Only the converters close to the fault (\acp{GFVSC} 1 and 2) activate the FVB-L strategy but not those far from it (\acp{GFVSC} 3 and 4). This behaviour is due to the logic rules of Fig.~\ref{fig:VSC_GF_FVB_L_logic} and an appropriate design of the FVB-L controller parameters. Therefore, this strategy is very effective whether implemented with CSA or HCL current limiters. Strategy FVB-WACS can provide a positive or/and negative supplementary voltage set-point during the first swing. This action depends on whether the converter frequencies are above or below the frequency of the COI (see Fig.~\ref{fig:Kundur_sim2_freq_COI}). Thus, \acp{GFVSC} 1 and 2 provide a positive supplementary voltage set-point $\Delta v_{f,i}^{ref,TS}$, slowing down the \acp{GFVSC}. While \acp{GFVSC} 3 and 4 provide a negative $\Delta v_{f,i}^{ref,TS}$, they accelerate as is shown in Fig.~\ref{fig:Kundur_sim2_bVf_TS} (blue and purple), reducing the risk of loss of synchronism. Strategy FVB-WACS is also effective when using CSA or HCL. 

\FloatBarrier

\begin{figure}[!htbp]
		\begin{center}
		\centering
		\includegraphics[width=0.65\columnwidth]{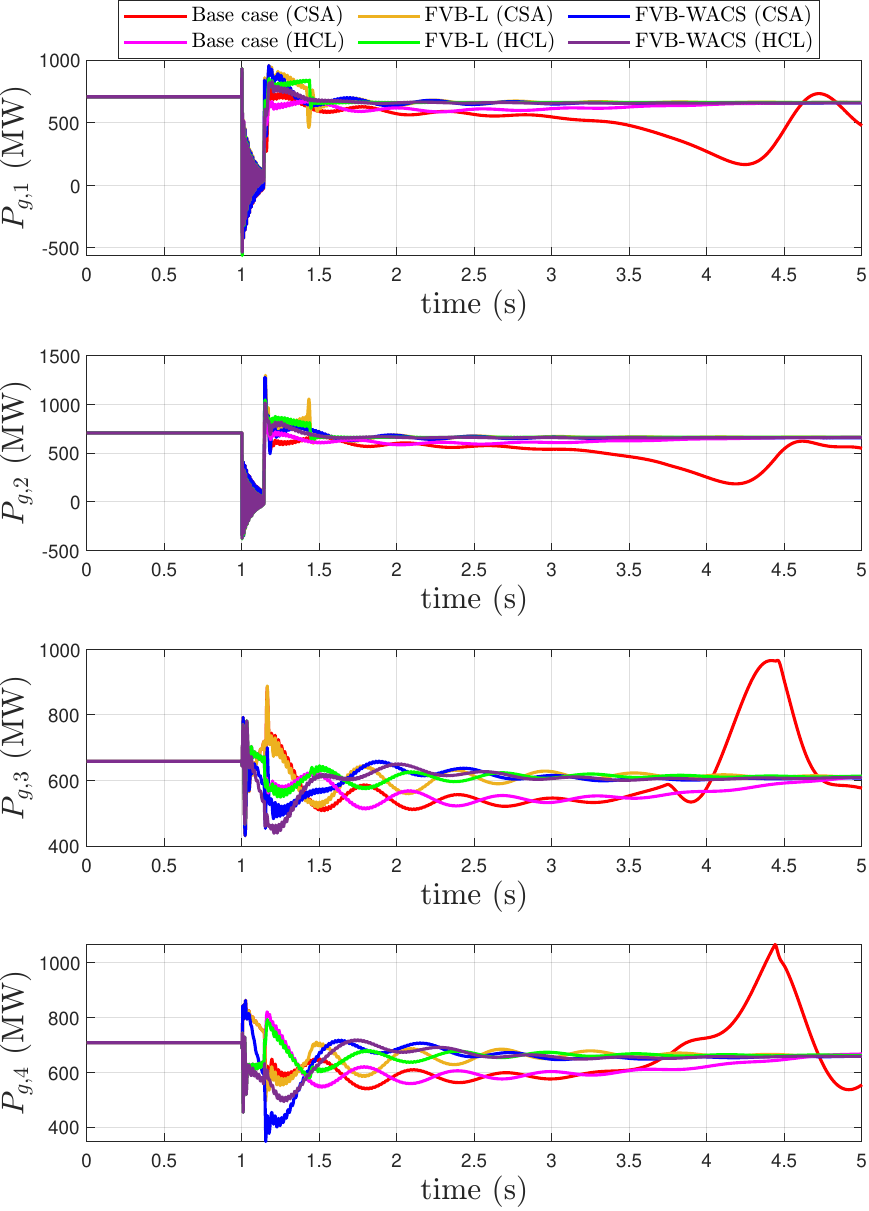}
		\vspace{-0.1cm}
		\caption{Fault I cleared after 140~ms. Active-power injections of the VSCs.}
		\label{fig:Kundur_sim2_Pg}
	\end{center}
\end{figure}

\begin{figure}[!htbp]
		\begin{center}
		\centering
		\vspace{-0.2cm}
		\includegraphics[width=0.65\columnwidth]{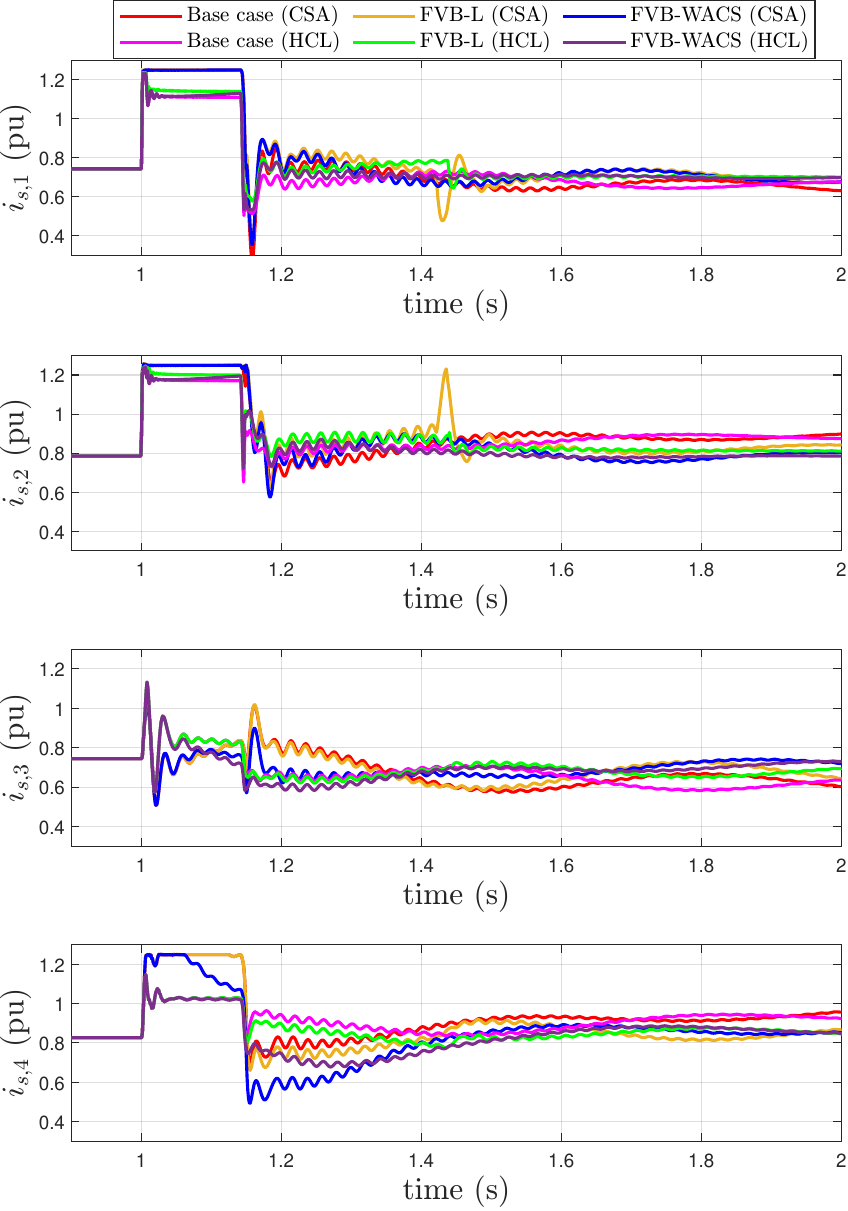}
		\vspace{-0.1cm}
		\setlength{\belowcaptionskip}{-10pt}
		\caption{Fault I cleared after 140~ms. Current injections of the VSCs.}
		\label{fig:Kundur_sim2_Ig}
	\end{center}
\end{figure}

Fig.~\ref{fig:Kundur_sim2_Pg} shows the active power injections of the \acp{GFVSC} and Fig.~\ref{fig:Kundur_sim2_Ig} their current injections. The fault activates the current limiters (either CSA or HCL) in those converters closer to it, leading to fast variations in the active-power injections until the fault clearing. After the fault is cleared, converters supply a positive or negative additional voltage set-point $\Delta v_{f,i}^{ref,TS}$, by the action of FVB controllers (FVB-L and FVB-WACS). The supplementary voltage set-points increase or decrease the active powers injections of the converters according to (\ref{eq:VSC_pg_electrical}). Precisely, (electrical) active-power injections, $p_{g,i}$, drive the slowing down or the acceleration of the \acp{GFVSC}, according to~(\ref{eq_VSC_V_VSM_v1}).
Current limiters CSA are activated just after the fault in \acp{GFVSC} 1, 2 and 3, where the current limit is reached ($i_{s,i}^{max}=1.25$). The current injections of those converters closer to the fault hit a peak, reaching their maximum allowed value according to Fig.~\ref{fig:VSC_V_CSA_diagram}. In the cases where the CSA is used alone (without FVBs), converters remain at the maximum limit value allowed by CSA until the fault clearance. Meanwhile, when HCLs are used, the current injections of \acp{GFVSC} initially reach the CSA limit, and then they are reduced by the virtual-impedance action, impeding the loss of synchronism and improving transient stability.

The effectiveness of FVBs when using HCLs is due to the fact that actions of both controllers take place at different stages of the transient. During the fault, the HCL has priority and it introduces a supplementary voltage set-point to the \ac{GFVSC}. Once the fault has been cleared, the HCL is disabled and the FVB introduces a supplementary voltage set-point to the \ac{GFVSC}. Therefore, the use of FVBs together with HCLs takes advantages of both.

\FloatBarrier
\vspace{-0.1cm}
\subsection{Critical clearing times (CCTs)}\label{sec:Results2_VSC_V_Kundur_CCTs1}
\noindent Tables~\ref{tab:Kundur_CCTs} and \ref{tab:Kundur_CCTs_HCL} show the critical clearing times (CCTs) of the faults described in Table~\ref{tab:Kundur_Faults_description} with/without FVBs, using CSA or HCL current limiters. Strategy FVB-WACS increases the CCTs of all faults. In contrast, the FVB-L strategy significantly increases the CCTs only on faults I and II, with no impact on faults III and IV. This behaviour is because the FVB-L strategy is designed to be activated only for severe-enough faults as a consequence of the activation conditions of the local strategy FVB-L described in Section~\ref{sec:VSC_V_TS_FVB_L}. Comparing Tables~\ref{tab:Kundur_CCTs_HCL} and \ref{tab:Kundur_CCTs}, results shows that HCL with no FVBs significantly improves the CCTs of faults III and IV with respect to CSA with no FVBs. However, the improvements for faults I and II are very small. This confirms that HCL is more effective than CSA.

\FloatBarrier
The use of HCL algorithms with FVB strategies improves the CCTs in all faults, even in those where the FVBs and the HCLs cannot improve when applied independently. The best results are obtained when implementing FVB-WACS with HCL current limiter.

\begin{table}[!htbp]
	\begin{center}
		\caption{Fault description.}
			\begin{tabular}{|l|c|c|c|}
				\hline
				          & Short circuit & close  & \multirow{2}{*}{clearing  }                                     \\
				          & at line $i-j$ & to bus &                                                \\ \hline
				Fault I   & 7-8a          & 7      & Disconnect 7-8a                                \\
				Fault II  & 5-6           & 5      & short circuit cleared  (line not disconnected) \\
				Fault III & 10-11         & 11     & short circuit cleared  (line not disconnected) \\
				Fault IV  & 8-9a          & 8      & Disconnect 8-9a                                \\
				\hline
			\end{tabular}
		\label{tab:Kundur_Faults_description}
		\caption{CCTs with CSA.  }
			\begin{tabular}{|l|c|c|ccc|}
				\hline
				\textbf{CCT}   & \multirow{2}{*}{\textbf{Base case}}  & \multirow{2}{*}{\textbf{FVB-L}  }  & \textbf{FVB-WACS}  &       with & delay        \\ 
				(ms)         &               &                   & $\tau=0$ ms & $50$ ms & $100$ ms \\ \hline
				Fault I      & 130           & 250               & 270         & 270     & 260      \\
				Fault II     & 270           & 310               & 360         & 340     & 320      \\
				Fault III    & 220           & 220               & 230         & 230     & 230      \\
				Fault IV     & 420           & 420               & 880         & 870     & 890      \\
				\hline
			\end{tabular}
		\label{tab:Kundur_CCTs}
		\caption{CCTs with HCL.}
			\begin{tabular}{|l|c|c|ccc|}
				\hline
				\textbf{CCT} & \textbf{Base case} & \textbf{FVB-L}   & \multicolumn{2}{c}{\textbf{FVB-WACS (HCL)}} & with delay \\ 
				(ms)         & \textbf{(HCL)} & \textbf{(HCL)}	   & $\tau=0$ ms & $50$ ms & $100$ ms \\ \hline
				Fault I      & 140           & 280               & 290         & 290     & 280      \\
				Fault II     & 280           & 330               & 370         & 370     & 360      \\
				Fault III    & 600           & 600               & 700         & 700     & 690      \\
				Fault IV     & 510           & 510               & 930         & 950     & 940      \\
				\hline
			\end{tabular}
		\label{tab:Kundur_CCTs_HCL}
	\end{center}
\end{table}

%\FloatBarrier
\subsection{Impact of communication latency}\label{sec:Results2_VSC_V_Kundur_latency}

\noindent The impact of communication latency on the performance of strategy FVB-WACS, using CSA and HCL current-limitation algorithms, has been analysed. The input error signal of FVB-WACS of Fig.~\ref{fig:VSC_GF_FVB_WACS} with a communication delay reads: 
\begin{equation}\label{eq:FB_WACS_w_ref_TS_delay}
	u_{i}=e^{-s \tau} (\omega_{COI}-\omega_{i})
\end{equation}
Total communication latencies ($\tau$) of 50~ms and 100~ms will be analysed, which are consistent with realistic delays in WACS~\cite{Chow2015}.

 CCTs obtained with strategy FVB-WACS with communication latency, when using CSA and HCL current limiters are shown in the last three columns of Table~\ref{tab:Kundur_CCTs_HCL}. The CCTs decrease as the communication delay increases. Results prove that strategy FVB-WACS is robust against communication latency, when using the two options of current-limitation strategies: CSA and HCL.

\FloatBarrier
\newpage
\section{Conclusions}\label{sec:conclusion}
This paper analysed the impact of hybrid current limiters (HCLs) and fast voltage boosters (FVBs) in \acp{GFVSC} on transient stability of power systems with 100\% converter-interfaced generation. 

The conclusions obtained in this chapter are as follows:
\begin{itemize}
    \item HCLs are effective in limiting the current of \acp{GFVSC} (their main application) and they can also improve transient stability, significantly, depending on the fault. 
    \item FVBs in \acp{GFVSC} improve transient stability when using CSA and HCL current limiters. Moreover, FVBs together with HCL significantly improve transient stability for all faults analyzed. This means that FVBs and HCLs are complementary and compatible.
    \item The effectiveness of FVBs when using HCLs is due to the fact that the actions of both controllers take place at different stages of the transient. During the fault, the HCL has priority and it introduces a supplementary voltage set-point to the \ac{GFVSC}. After the fault clearing, the HCL does not act any more and then FVB introduces a supplementary voltage set-point to the \ac{GFVSC}. Hence, the use of FVBs together with HCLs takes advantages of both.
	\item FVB strategies improve transient stability with local (FVB-L) and global (FVB-WACS) measurements. FVB-WACS is the most effective, when using both current-limitation algorithms analysed (CSA and HCL).
    \item FVB-WACS is robust against communication latency, when using both current-limitation algorithms analysed (CSA and HCL). 
\end{itemize}

\vspace{-0.2cm}
\section*{Appendix: data}
\noindent Table~\ref{tab:VSC_parameters} depicts the data of the grid-forming VSCs. The data of the original two-area Kundur's test system can be found in~\cite{Kundur1994a}.
 In this work the same conditions of~\cite{RAvilaM2022} were considered: (Load at bus 7: 917 MW \& 100 MVAr; load at bus 9: 1817 MW \& 100 MVAr). Nominal frequency is 50 Hz. 

\FloatBarrier
\begin{table}[!htbp]
	\caption{Parameters of the VSCs}
	\begin{center}
		\scalebox{0.8}{
			\begin{tabular}{|l|c|}
				\hline
				\textbf{Parameters}                                                                          &                                         \\
				VSC's rating are base values for pu                                                          &                                         \\
				\hline
				Rating VSC, DC voltage, AC voltage                                                           & 900 MVA, $640$ kV, $300$ kV         \\
				Max. modulation index ($m_{i}^{max} = \sqrt{\frac{3}{2}} \cdot \frac{V_{dc,B}}{2 V_{ac,B}}$) & 1.31 pu                                 \\
				Series filter resistance ($r_{f,i}$)/reactance ($x_{f,i}$)                                   & 0.005 pu / 0.15 pu                      \\
				Shunt filter capacitance ($C_{f,i}$)                                                         & 0.0660 pu                               \\
				Transformer resistance ($r_{c,i}$)/reactance ($x_{c,i}$)                                     & 0.005 pu / 0.15 pu                      \\
				(900 MVA 300/220 kV transformer)                                                             &                                         \\
				Current prop./int. control  ($K_{C,P,i}$/$K_{C,I,i}$)                                          & 0.73 pu / 1.19 pu/s                     \\
				Voltage prop./int. control   ($K_{V,P,i}$/$K_{V,I,i}$)                                         & 0.52 pu / 1.16 pu/s                     \\
				Virtual transient resistance   ($r_{V,i}$/$T_{VR,i}$)                                        & 0.09 pu / 0.0167 s                      \\
				Emulated inertia ($H_{VSC,i}$) of VSCs 1, 2, 3 \& 4                                              & 4.5  s / 4.5 s / 4.175 s / 6.175 s                         \\
				Primary freq. controller gain. ($D_{VSC,i}$)                                                 & 20 pu                                   \\
				\hline
			\end{tabular}
		}
		\label{tab:VSC_parameters}
	\end{center}
\end{table}

Parameters of the FVBs:
\begin{itemize}
	\item FVB-L: Figs.~\ref{fig:VSC_GF_FVB_L}-\ref{fig:VSC_GF_FVB_L_logic} with: $v_{A,i}=0.5$~pu, $v_{B,i}=0.9$~pu, $\omega_{thres,i}=10^{-3}$~pu, $\Delta v_{f,i}^{max}=0.15$~pu.
	\item FVB-WACS: Fig.~\ref{fig:VSC_GF_FVB_WACS} with: $K_{FVB,i}=50$~pu, $T_{f,i}=0.1$~s, $T_{W,i}=10$~s, $\Delta v_{f,i}^{max}=0.15$~pu and $\epsilon_{i}=10^{-3}$~pu.
\end{itemize}
Parameters of the current limiters:
\begin{itemize}
    \item CSA: $i_{s,i}^{max}=1.25$~pu (equal priority for $d-q$ axes).
    \item VI-CL: $i_{VI,i}^{max}=1.0$~pu, $k_{p_{r_{VI}}} = 0.098$~pu, $\sigma_{x/r}=5$.
    \item  HCL: Parameters of CSA and VI-CL are used.  
\end{itemize}

\section*{Acknowledgements}\label{sec:agradecimientos}
\noindent Work supported by the Spanish Government under a research project ref. PRE2019-088084 and RETOS Project Ref. RTI2018-098865-B-C31 (MCI/AEI/FEDER, UE); and by Madrid Regional Government under PROMINT-CM Project Ref. S2018/EMT-4366. Accepted in the 23rd Power Systems Computation Conference (PSCC 2024).

Contact: \{regulo.avila, luis.rouco, aurelio, lukas.sigrist\}@iit.comillas.edu, javier.renedo@ieee.org, \\ xavier.guillaud@centralelille.fr, taoufik.qoria@gmail.com.

%\newpage
%\printbibliography
%\FloatBarrier
% \bibliographystyle{IEEEtran}
%\vspace{-0.2cm}
% \bibliography{biblio_VSC_GFo_red_v1}

\begin{thebibliography}{10}
\providecommand{\url}[1]{#1}
\csname url@samestyle\endcsname
\providecommand{\newblock}{\relax}
\providecommand{\bibinfo}[2]{#2}
\providecommand{\BIBentrySTDinterwordspacing}{\spaceskip=0pt\relax}
\providecommand{\BIBentryALTinterwordstretchfactor}{4}
\providecommand{\BIBentryALTinterwordspacing}{\spaceskip=\fontdimen2\font plus
\BIBentryALTinterwordstretchfactor\fontdimen3\font minus \fontdimen4\font\relax}
\providecommand{\BIBforeignlanguage}[2]{{%
\expandafter\ifx\csname l@#1\endcsname\relax
\typeout{** WARNING: IEEEtran.bst: No hyphenation pattern has been}%
\typeout{** loaded for the language `#1'. Using the pattern for}%
\typeout{** the default language instead.}%
\else
\language=\csname l@#1\endcsname
\fi
#2}}
\providecommand{\BIBdecl}{\relax}
\BIBdecl

\bibitem{Paolone2020}
M.~Paolone, T.~Gaunt, X.~Guillaud, M.~Liserre, S.~Meliopoulos, A.~Monti, T.~{Van Cutsen}, V.~Vittal, and C.~Vournas, ``{Fundamentals of power systems modelling in the presence of converter-interfaced generation},'' \emph{Electric Power Systems Research}, vol. 189, no. 106811, pp. 1--33, 2020.

\bibitem{Barker2021}
C.~Barker, A.~Adamczyk, J.~Fradley, and O.~Jasim, ``{Providing Synchronous Grid Forming Capability through HVDC Transmission},'' in \emph{Proc. 17th IET Conference on AC and DC Power Transmission (ACDC)}, Dec. 2021, pp. 161--166.

\bibitem{DArco2015}
S.~{D'Arco}, J.~A. Suul, and O.~B. Fosso, ``{A Virtual Synchronous Machine implementation for distributed control of power converters in SmartGrids},'' \emph{Electric Power Systems Research}, vol. 122, pp. 180--197, 2015.

\bibitem{jroldan2019}
J.~Rold\'an-P\'erez, A.~Rodr\'iguez-Cabero, and M.~Prodanovic, ``{Design and analysis of virtual synchronous machines in inductive and resistive weak grids},'' \emph{IEEE Trans. on Energy Conversion}, vol.~34, no.~2, pp. 1818--1828, 2019.

\bibitem{Qoria2020}
T.~Qoria, E.~Rokrok, A.~Bruyere, B.~Francois, and X.~Guillaud, ``{A PLL-Free Grid-Forming Control With Decoupled Functionalities for High-Power Transmission System Applications},'' \emph{IEEE Access}, no. 106765, pp. 197\,363--197\,378, 2020.

\bibitem{Choopani2020}
M.~Choopani, S.~H. Hosseinian, and B.~Vahidi, ``{New Transient Stability and LVRT Improvement of Multi-VSG Grids Using the Frequency of the Center of Inertia},'' \emph{IEEE Trans. on Power Systems}, vol.~35, no.~1, pp. 527--538, 2020.

\bibitem{Xiong2021}
X.~Xiong, C.~Wu, P.~Cheng, and F.~Blaabjerg, ``{An Optimal Damping Design of Virtual Synchronous Generators for Transient Stability Enhancement},'' \emph{IEEE Trans. on Power Electronics}, vol. doi: 10.1109/TPEL.2021.3074027, pp. 1--5, 2021.

\bibitem{Collados2023}
C.~Collados-Rodriguez, D.~{Westerman Spier}, M.~Cheah-Mane, E.~Prieto-Araujo, and O.~Gomis-Bellmunt, ``{Preventing loss of synchronism of droop-based grid-forming converters during frequency excursions },'' \emph{International Journal of Electrical Power \& Energy Systems}, vol. 148, no. 108989, pp. 1--8, 2023.

\bibitem{XiongX_2021a}
\BIBentryALTinterwordspacing
X.~Xiong, C.~Wu, and F.~Blaabjerg, ``\BIBforeignlanguage{en}{An {Improved} {Synchronization} {Stability} {Method} of {Virtual} {Synchronous} {Generators} {Based} on {Frequency} {Feedforward} on {Reactive} {Power} {Control} {Loop}},'' \emph{\BIBforeignlanguage{en}{IEEE Transactions on Power Electronics}}, vol.~36, no.~8, pp. 9136--9148, Aug. 2021. [Online]. Available: \url{https://ieeexplore.ieee.org/document/9328607/}
\BIBentrySTDinterwordspacing

\bibitem{RAvilaM2022}
R.~E. \'Avila-Mart\'inez, J.~Renedo, L.~Rouco, A.~Garc\'ia-Cerrada, L.~Sigrist, T.~Qoria, and X.~Guillaud, ``{Fast voltage boosters to improve transient stability of power systems with 100\% of grid-forming VSC-based generation},'' \emph{IEEE Trans. on Energy Conversion}, vol.~37, no.~4, pp. 2777--2789, 2022.

\bibitem{BlaabjergFTSAngle2022}
M.~Chen, D.~Zhou, and F.~Blaabjerg, ``Enhanced {Transient} {Angle} {Stability} {Control} of {Grid}-{Forming} {Converter} {Based} on {Virtual} {Synchronous} {Generator},'' \emph{IEEE Trans. on Industrial Electronics}, vol.~69, no.~9, pp. 9133--9144, Sep. 2022.

\bibitem{SiW2023}
\BIBentryALTinterwordspacing
W.~Si and J.~Fang, ``\BIBforeignlanguage{en}{Transient {Stability} {Improvement} of {Grid}-{Forming} {Converters} {Through} {Voltage} {Amplitude} {Regulation} and {Reactive} {Power} {Injection}},'' \emph{\BIBforeignlanguage{en}{IEEE Transactions on Power Electronics}}, vol.~38, no.~10, pp. 12\,116--12\,125, Oct. 2023. [Online]. Available: \url{https://ieeexplore.ieee.org/document/10167833/}
\BIBentrySTDinterwordspacing

\bibitem{BoFang}
B.~Fan, T.~Liu, F.~Zhao, H.~Wu, and X.~Wang, ``A review of current-limiting control of grid-forming inverters under symmetrical disturbances,'' \emph{IEEE Open Journal of Power Electronics}, vol.~3, no. xxx, pp. 955--969, 2022.

\bibitem{Qoria_VSC_current_limit2020}
T.~Qoria, F.~Gruson, F.~Colas, X.~Kestelyn, and X.~Guillaud, ``{Current limiting algorithms and transient stability analysis of grid-forming VSCs},'' \emph{Electric Power Systems Research}, vol. 189, no. 106726, pp. 1--8, 2020.

\bibitem{Qoria_VSC_CCT2020}
T.~Qoria, F.~Gruson, F.~Colas, G.~Denis, T.~Prevost, and X.~Guillaud, ``{Critical Clearing Time Determination and Enhancement of Grid-Forming Converters Embedding Virtual Impedance as Current Limitation Algorithm},'' \emph{IEEE Journal of Emerging and Selected Topics in Power Electronics}, vol.~8, no.~2, pp. 1050--1061, 2020.

\bibitem{Rokrok_TS2021}
E.~Rokrok, T.~Qoria, A.~Bruyere, B.~Francois, and X.~Guillaud, ``{Transient Stability Assessment and Enhancement of Grid-Forming Converters Embedding Current Reference Saturation as Current Limiting Strategy},'' \emph{IEEE Trans. Power Systems}, vol. Online: doi: 10.1109/TPWRS.2021.3107959, pp. 1--12, 2021.

\bibitem{Guangya2021}
\BIBentryALTinterwordspacing
K.~Vatta~Kkuni and Y.~Guangya, ``Effects of current limit for grid forming converters on transient stability: analysis and solution,'' 06 2021. [Online]. Available: \url{https://ssrn.com/abstract=4356280 or http://dx.doi.org/10.2139/ssrn.4356280}
\BIBentrySTDinterwordspacing

\bibitem{Laba2023}
Y.~Laba, A.~Bruyere, F.~Colas, and X.~Guillaud, ``{Virtual Power-Based Technique for Enhancing the Large Voltage Disturbance Stability of HV Grid-Forming Converters},'' \emph{TechRxiv}, vol. Online, pp. 1--8, Aug. 2023.

\bibitem{Kundur1994a}
P.~Kundur, \emph{Power System Stability and Control}.\hskip 1em plus 0.5em minus 0.4em\relax McGraw Hill, 1994.

\bibitem{Paquette}
A.~D. Paquette and D.~M. Divan, ``Virtual impedance current limiting for inverters in microgrids with synchronous generators,'' \emph{IEEE Trans. on Industry Applications}, vol.~51, no.~2, pp. 1630--1638, 2015.

\bibitem{LuisDM2017}
L.~Diez-Maroto, L.~Vanfretti, M.~S. Almas, G.~M. J\'{o}nsd\'{o}ttir, and L.~Rouco, ``{A WACS exploiting generator Excitation Boosters for power system transient stability enhancement},'' \emph{Electric Power Systems Research}, vol. 148, pp. 245--253, 2017.

\bibitem{LuisDM2019}
L.~D\'{i}ez-Maroto, J.~Renedo, L.~Rouco, and F.~Fern\'{a}ndez-Bernal, ``{Lyapunov Stability Based Wide Area Control Systems for Excitation Boosters in Synchronous Generators},'' \emph{IEEE Trans. on Power Systems}, vol.~34, no.~1, pp. 194--204, 2019.

\bibitem{L2EP_VSC_GF_2020}
L2EP-LILLE, ``{VSC\_Lib: Grid Forming Models for Matlab/SimPowerSystem},'' vol. https://github.com/l2ep-epmlab/ (accessed 08-07-2020), 2020.

\bibitem{MIGRATE_WP3_2018a}
T.~Qoria, Q.~Cossart, C.~Li, X.~Guillaud, F.~Colas, F.~Gruson, and X.~Kestelyn, ``{WP3-Control and Operation of a Grid with 100\% Converter-Based Devices. D3.2: Local control and simulation tools for large transmission systems},'' MIGRATE Project, Tech. Rep., 2018.

\bibitem{Qoria2019a}
T.~Qoria, F.~Gruson, F.~Colas, G.~Denis, T.~Prevost, and X.~Guillaud, ``{Inertia effect and load sharing capability of grid forming converters connected to a transmission grid},'' in \emph{15th IET Int. Conf. on AC and DC Power Transmission (ACDC), Coventry, UK}, 2019, pp. 1--6.

\bibitem{Chow2015}
F.~Zhang, Y.~Sun, L.~Cheng, X.~Li, J.~H. Chow, and W.~Zhao, ``{Measurement and Modeling of Delays in Wide-Area Closed-Loop Control Systems},'' \emph{IEEE Trans. on Power Systems}, vol.~30, no.~1, pp. 2426--2433, 2015.

\end{thebibliography}

\end{document}